\documentclass[prb,twocolumn,aps,showpacs]{revtex4}
\usepackage{graphicx}
\def\ltsim{\vbox {\hbox{\lower .8\baselineskip \hbox{$<$}} \break
                 \hbox{\lower 0.2\baselineskip \hbox{$\sim$}} } }

\begin{document}

\title{Decoherence of Einstein-Podolsky-Rosen  pairs in a noisy Andreev entangler}
\author{\' Emilie Dupont and Karyn Le Hur}
\affiliation{D\'epartement de Physique, Universit\'e de Sherbrooke, 
Sherbrooke, Qu\'ebec, Canada J1K 2R1}

\begin{abstract}
We investigate quantum noise effect on the transportation of nonlocal Cooper pairs accross the
realistic Andreev entangler which consists of an s-wave superconductor coupled to two small
quantum dots at resonance which themselves are coupled to normal leads. The noise emerges due to voltage fluctuations felt by the electrons residing on the two dots as a result of the finite resistances in the gate leads or of any resistive lead capacitively coupled to the dots. In the ideal noiseless case, the setup provides a trustable source of mobile and nonlocal spin-entangled electrons and the transport is dominated by a two-particle Breit-Wigner resonance that allows the injection of two spin-entangled electrons  into different leads at the same energy [P. Recher, E. V. Sukhorukov, and D. Loss, Phys. Rev. B {\bf 63}, 165314 (2001)]. We seek to revisit the transport of those nonlocal Cooper pairs as well as the efficiency of such an Andreev entangler when including the quantum noise (decoherence).
\end{abstract}

\date{\today} 
\maketitle

The concept of nonlocal pairwise-entangled quantum states\cite{Schrodinger}, namely the Einstein-Podolsky-Rosen (EPR) pairs\cite{Einstein} is extremely fundamental for testing the violation of Bell inequality\cite{bell} and at the same time is also crucial for efficient quantum communication\cite{B} and quantum teleportation\cite{BB} for example. Tests on the entanglement of massless particles like the photons already exist\cite{Aspect}
but not yet for massive particles such as the electrons. A concrete challenge is to realize an entangler of electrons, {\em i.e.}, a device that generates spin singlets that are made out of two electrons. Recently, an interesting setup involving an s-wave superconductor weakly-coupled to two quantum dots, which themselves are coupled to normal leads has been envisioned in Ref. \onlinecite{Recher}. The spin correlations of an s-wave superconductor induce a spin-singlet state between two electrons, each of which can now reside on a separate quantum dot; this results in the formation of a nonlocal Cooper pair described by the quantum state
\begin{equation}
|DD\rangle=[d^{\dagger}_{1\uparrow}d^{\dagger}_{2\downarrow}-d^{\dagger}_{1\downarrow}d_{2\uparrow}^{\dagger}]|i\rangle,
\end{equation}
where $d^{\dagger}_{l\sigma}$ produces an electron with spin $\sigma$ on the dot $l$ and $|i\rangle=|0\rangle_S |0\rangle_D|\mu_l\rangle$ ($|0\rangle_S$ is the quasiparticle vacuum for the
superconductor, $|0\rangle_D$ means that both dot levels $\epsilon_l$ are unoccupied, and $|\mu_l\rangle$ defines the occupation of the leads which are filled with electrons up to the electrochemical potential $\mu_l$) is the initial state. A prerequisite to perform such a nonlocal spin-entangled electron state is the Coulomb blockade phenomenon which typically forbids double occupancy on each quantum dot: the spin singlet coming from the Cooper pair remains preserved in this process even though the two involved partners are well separated physically because they reside on different quantum dots. More precisely, such an Andreev entangler\cite{Andreev} of EPR pairs has been predicted to be well efficient when\cite{Recher} ${\cal E}/\gamma\gg (k_F \delta r)$ where ${\cal E}^{-1}=1/(\pi\Delta) + 1/U$, $U$ being twice the charging energy of a given dot, $\Delta$ the superconducting gap, $\delta r=|\delta {\bf r}|$ which might be of the order of the distance between dots denotes the distance between the points on the superconductor from which electrons 1 and 2 tunnel into the dots, ${\bf k}_F$ $(k_F=|{\bf k}_F|)$ the Fermi momentum in the superconductor, and $\gamma$ the effective tunneling rate between the dots and the normal leads. We have assumed a two-dimensional
superconductor. The parasitic direct Andreev process where two spin-entangled electrons from the superconductor navigate accross the same dot can be minimized due to the Coulomb blockade effect. In principle, assuming that the tunnel barriers between the dots and the superconductor are weak enough, this allows to suppress the other parasitic quantum process namely the superconducting cotunneling\cite{Sautet} (transfer of an electron from one dot to another via the superconductor). In this setup, there is a direct correspondance between the idea of preserving the spin entanglement on the two dots - implying that the distance between the two dots is smaller than the coherence length $\xi$ of the superconductor - and the result to get a two-particle Breit-Wigner resonance allowing the spin injection of two spin-entangled electrons into different leads at exactly the same orbital energy and hence a quite prominent stationary current when a bias voltage is applied between the superconductor and the two leads.  Remember the one to one correspondance between non-local spin-entanglement on the two dots and prominent stationary current between the superconductor and the leads triggered by the crossed Andreev reflection. A similar conclusion arises when replacing the single-level quantum dots by one-dimensional Luttinger liquids\cite{Recher2,Smitha} however the efficiency condition of the entangler is modified as\cite{Recher3} $(\Delta/2\mu)^{2\gamma_{\rho}}\gg (k_F\delta r)$ where $\mu$ is the difference of electrochemical potentials between the superconductor and the Luttinger leads and $\gamma_{\rho}=(K_{\rho}+K_{\rho}^{-1})/4 -1/2>0$, where $K_{\rho}<1$ is the Luttinger parameter,
stands for the exponent for tunneling into the bulk of Luttinger leads. 

The nonlocal spin entanglement might be detected through (current) noise measurements\cite{Burkard}. In fact, using a beam splitter in the setup of Ref. \onlinecite{Recher}, Samuelsson, Sukhorukov, and B\"{u}ttiker have shown
that the electrical current noise could serve as a test of spin-entanglement\cite{Samuelsson}. Following Ref. \onlinecite{Blatter}, a diagnosis on spin current noise and Bell inequalities in the same setting has been done by 

\begin{figure}[ht]
\begin{picture}(235,220)
\leavevmode\centering\includegraphics{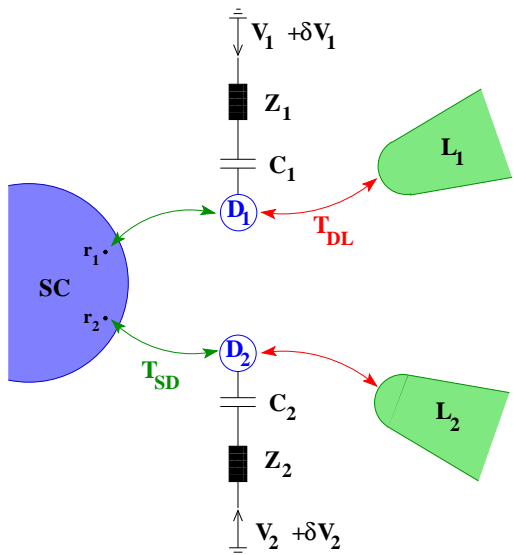}
 
 \end{picture}
\caption{(Color online) The noisy entangler: Two spin-entangled electrons forming a Cooper pair tunnel with amplitude $T_{SD}$ from points ${\bf r}_1$ and ${\bf r}_2$ of the superconductor (SC) to two dots
$D_1$ and $D_2$ by means of crossed Andreev reflection. The dots are coupled to normal leads $L_1$ and $L_2$ with tunneling amplitude $|T_{DL}|>|T_{SD}|$. The SC and leads are maintained at chemical potentials $\mu_S$ and $\mu_l$ respectively so that $\mu=\mu_S-\mu_l>0$. 
We show the dissipative leads (which are not necessarily the gate leads) embodied by the impedances $Z_1$ and $Z_2$ producing voltage fluctuations $(\delta V_l(t))$ on each dot.} 
\label{setup}
\end{figure}

\hskip -0.3cm Sauret, Martin, and Feinberg\cite{Sauret}. Bell-inequality checks
in solid-state systems can be thought as interesting generalizations of the corresponding tests with photons\cite{Kawabata,Beenakker}.  

Many theoretical papers are devoted to the crossed Andreev reflection (the elastic emission of spin-entangled electrons in different leads) as well as to the distance-contribution of the crossed Andreev reflection\cite{Lesovik} and to its implications on the conductance matrix\cite{Byers,Martin,Feinberg,Hekking} or on the current-current cross correlation\cite{Pistolesi}.

On the other hand, the destruction of quantum-mechanical phase coherence due to coupling of a system to a dissipative bath is a subject of great importance partly because of its connection to fundamental
issues related to the quantum measurement process or the quantum-classical crossover. 
Some specific condensed matter setups have been also proposed to answer the question whether decoherence at zero temperature is possible at all \cite{Markus,Marquardt1,Marquardt2}. 
Following this line of ideas, it is then interesting to ask to what extent zero-point fluctuations of the electrical environments of the two dots will affect the non-local Cooper pair of charge $2e$ and hence the stationary current(s) discussed in Ref. \onlinecite{Recher}.  
Therefore, we now turn our attention to the specific setup of Fig. 1 where the external impedances $Z_{1}(\omega)$ and $Z_2(\omega)$ show resistive behavior at low frequencies. Albeit voltage fluctuations from gate leads in GaAs, in double dot geometries, are thought to be quite small\cite{Viel} and often ignored, the quantum noise may emerge from any resistive source capacitively coupled to the dots. More precisely, we have in mind the GaAs heterostructures of Ref. \onlinecite{Rimberg} where quantum dots are built in the proximity of two-dimensional electron gases with (tunable) resistances which can even reach few $k\Omega$ when exploring the low density limit of those two-dimensional electron gases. The noise might also stem from
the metallic leads L1 and L2 on Fig. 1; indeed, similar results can be derived. The quantum noise will couple to the nonlocal Cooper pair through the requirement that to hop on the dot an electron must transmit a part of its energy to the electrical environment (bath) and thus the $|DD\rangle$ state must be thought as:
\begin{equation}
|DD\rangle |D_B\rangle = e^{-i\delta\phi}[d^{\dagger}_{1\uparrow}d^{\dagger}_{2\downarrow}-d^{\dagger}_{1\downarrow}d_{2\uparrow}^{\dagger}]|i\rangle|i_B\rangle,
\end{equation}
describing explicitly the ``entanglement'' of the nonlocal Cooper pair to the quantum noise; 
$|i_B\rangle$ embodies the initial bath (ground) state,  $|D_B\rangle = e^{-i\delta\phi} |i_B\rangle$ describes the bath excited state when the two electrons reside on their respective dots and at a general level $\delta\phi(t)=(e/\hbar)\int_0^t (\delta V_{1}(t') +\delta V_{2}(t'))dt'$ embodies the phase conjugate to the sum of voltage fluctuations felt by the two dots\cite{Nazarov}. Note that Eq. (2) stems from the fact that in the presence of voltage fluctuations, the dot's electron annihilation operators will be modified as\cite{Marquardt2} $d_{l\alpha}\rightarrow d_{l\alpha}e^{i(e/\hbar)\int_0^t \delta V_{l}(t') dt'}$. As a matter of fact, in the case of moderate ohmic resistors $R_1$ and $R_2$ such as 
$Z_l(\omega)=R_l$, $\langle e^{-i\delta\phi(t)}\rangle = \langle i_B| e^{-i\delta\phi(t)} |i_B\rangle$ always goes to zero at sufficiently long times both in the finite temperature regime and in the quantum realm\cite{Marquardt1} (at finite temperatures, one gets a relatively fast exponential decay versus time whereas approaching the quantum limit one reaches a less severe ``power-law'' decay). By averaging over the bath degrees of freedom, we can rewrite Eq. (2) as $|DD\rangle \langle i_B|D_B\rangle = \langle e^{-i\delta\phi} \rangle [d^{\dagger}_{1\uparrow}d^{\dagger}_{2\downarrow}-d^{\dagger}_{1\downarrow}d_{2\uparrow}^{\dagger}]|i\rangle$, hence we infer that the environments will contribute to the destruction of the (nonlocal) Cooper pair object in the long-time limit. Our interest is to study the consequence of the nonlocal Cooper pair long-time ``decoherence'' on the current between the superconductor and the leads. By decoherence we explicitly refer to the long-time decay of the noisy Cooper pair through the factor $\langle e^{-i\delta\phi(t)} \rangle$ even though the spin entanglement is not destroyed. 

Even though electron spins in semiconductor heterostructures are in principle very robust to decoherence with dephasing time approaching microseconds\cite{Asc}, in the realistic setup of Ref. \onlinecite{Recher} one can nevertheless discuss the phenomenon of decoherence (decay) of a nonlocal charged-2e Cooper pair during the transportation process due to the electrical circuits existing in the vicinity of the dots; this is the subject which will be addressed in this paper. 
Our paper is organized as follows. In Sec. I, we introduce the model of the noisy Andreev entangler and subsequently we write a general formula for the current resorting to the T-matrix approach. In Sec. II, we discuss the result for  a single environment --- assuming for example that $R_2=0$ --- both in the limit of large resistances $R_1/R_K\gg 1$, $R_K=h/e^2=25.8k\Omega$ being typically the quantum of resistance, and in the maybe more realistic situation of moderate resistances $R_1/R_K\ll 1$. In Sec. III, we extend our results for the two bath situation where both $R_1$ and $R_2$ are assumed to be non-negligible.
In Sec. IV, we also take into account the parasitic direct Andreev processes where the two electrons issued from the superconductor eventually jump onto the same dot; this processes are nevertheless
reduced at low temperatures due to the prolific combination of the Coulomb blockade principle which typically forbids two electrons to jump simultaneously onto the same dot and the superconducting gap which avoids quasiparticle formation in the superconductor. In the conclusion, we seek to revisit the condition of efficiency for such a noisy entangler. Appendices are devoted to the details of calculations.

\section{Model of the noisy entangler}

\subsection{Hamiltonian of the noisy Andreev entangler}

Similar to Ref. \onlinecite{Recher}, we exploit a tunneling Hamiltonian description of the system, $H=H_0+H_T+H^{bath}_{D}$, where 
\begin{equation}
H_0=H_S+\sum_{l} (H_{D_l}+H_{L_l}) +\sum_l H^{bath}_l.
\end{equation}
The superconductor is embodied by the usual BCS theory\cite{Schrieffer} which for convenience has been summarized in Appendix A. In brief, $H_S=\sum_{k\sigma} E_{\bf k} \gamma^{\dagger}_{{\bf k}\sigma} \gamma_{{\bf k}\sigma}$ where $\sigma=\uparrow,\downarrow$ represents the spin index, 
$\gamma_{{\bf k}\sigma}$ describe excitations out of the BCS ground state $|0\rangle_S$ defined by
$\gamma_{{\bf k}\sigma}|0\rangle_S=0$, $E_{\bf k}=\sqrt{\Delta^2+\xi_{\bf k}^2}$ is the quasiparticle energy, and $\xi_{\bf k}=\epsilon_{\bf k}-\mu_S$ is the normal state single-electron energy counted from the level
$\mu_S$ where live the Cooper spin-singlet particles. Both dots are embodied by a single level
with energy $\epsilon_l$ very close to $\mu_S$ and are typically governed by an Anderson model $H_{D_l}=\epsilon_l\sum_{\sigma} d^{\dagger}_{l\sigma}d_{l\sigma}+Un_{\uparrow}n_{\downarrow}$ and $l=1,2$. The resonant dot level $\epsilon_l$ can be adjusted by the related gate voltage (or by $V_l$ on Fig. 1). Other levels do not participate in the transport when the level spacing of the dots is sufficiently large implying $\delta\epsilon>\mu>k_B T$. Again, $\mu=\mu_S-\mu_l$ is the difference of electrochemical potentials between the superconductor and the leads. Moreover, through the on-site Coulomb $U$ repulsion a double occupied state is rather hindered to form on each dot; $U$ is equal to $2E_c$ where $E_c=e^2/(2C)$ is typically the charging energy on each dot and $C$ denotes the total dot's capacitance. Keep in mind that this Coulomb blockade argument stands for a key point in the efficiency of this Andreev entangler of EPR pairs\cite{Recher}. The leads are normal and embodied by a non-interacting theory $H_{L_l}=\sum_{{\bf k}\sigma} \epsilon_{\bf k} a^{\dagger}_{l{\bf k}\sigma} a_{l{\bf k}\sigma}$. We have to consider final two-particle
states of the form 
$|f \rangle = (1/\sqrt{2}) [ a^{\dagger}_{1{\bf p}\uparrow} a^{\dagger}_{2{\bf q}\downarrow} - a^{\dagger}_{1{\bf p}\downarrow} a^{\dagger}_{2{\bf q}\uparrow}]|i\rangle$ with energy $\epsilon_f=\epsilon_{\bf p}+\epsilon_{\bf q}$. 
The preserved spin singlet state is formed out of two electrons,
\begin{figure}[h]
\begin{picture}(60,60)
\leavevmode\centering\includegraphics{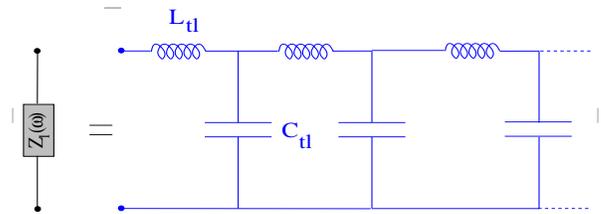}
 
 \end{picture}
 \vskip 0.2cm
\caption{(Color online) Impedances as transmission lines.} 
\label{setup}
\end{figure}
one being in the ${\bf p}$ state in lead 1 while the other one
is in the ${\bf q}$ state in lead 2. Since the total spin is conserved, the singlet state of the initial Cooper
pair will be conserved in the transport process and the final state must satisfy $S_z=0$. The $S_z=0$ configuration of the triplet state is excluded as long as the distance between the two dots is smaller
than $\xi$. Tunneling events from the dot $l$ to the lead $l$ or to the point ${\bf r}_l$ in the s-wave superconductor is described by the tunnel Hamiltonian $H_T=H_{SD}+(H_{DL}+H_{DL}^*)$, with
\begin{eqnarray}
H_{SD} &=& \sum_{l\sigma} T_{SD} d^{\dagger}_{l\sigma} \psi_{\sigma}({\bf r}_l)+h.c. = \sum_ l H_{SD_l},\\ \nonumber
H_{DL} &=& \sum_{l{\bf k}\sigma} T_{DL} a^{\dagger}_{l {\bf k} \sigma} d_{l \sigma} = \sum_l H_{D_l L_l}.
\end{eqnarray}
Here, $\psi_{\sigma}({\bf r}_l)$ annihilates an electron in the superconductor at the site ${\bf r}_l$ and 
$d^{\dagger}_{l\sigma}$ creates it again on dot $l$ 
with amplitude $T_{SD}$. Tunneling from the dot (lead) $l$ to the lead (dot) $l$ is described by the tunneling amplitude $T_{DL}$ $(T_{DL}^*)$. The ${\bf k}$ dependence of $T_{DL}$ can be safely neglected. Moreover, like in Ref. \onlinecite{Recher}, we
require the dot-lead coupling to be much stronger than the superconductor-dot coupling, {\em i.e.}, $|T_{SD}|<|T_{DL}|$, so that electrons that enter the dots from the superconductor will leave the quantum dots to the leads much
faster that new electrons can be provided to the dots from the superconductor. Additionally, a stationary
occupation due to the couplings to the
 leads is exponentially small if $\mu>k_B T$. Thus, in the asymmetric barrier case, the resonant dot levels $\epsilon_l$ are occupied only during a virtual process. Other
important parameters are the tunneling rates $\gamma_l=2\pi\nu_l|T_{DL}|^2$ and $\gamma_S=2\pi \nu_S |T_{DS}|^2$ where $\nu_l$ is the density of states per spin of the leads at the Fermi energy and $\nu_S$ for the superconductor will be defined as $1/\Delta$. Recall that we will work in the regime where $\gamma_l>\gamma_S$ and $\Delta,U,\delta\epsilon \gg \mu>\gamma_l,k_BT$ and close to the resonant condition for the dots $\epsilon_l\approx \mu_S$.

We model the impedance $Z_l(\omega)$ in a microscopic
fashion through a long dissipative transmission line composed of an 
infinite collection of
$L_{tl}C_{tl}$ oscillators (Fig. 2). 
Our environments are modeled in a usual way akin to 
Refs. \onlinecite{Markus} and \onlinecite{karyn}: the charge operator $\hat{Q}_{nl}$ on the capacitor between two inductances
$L_{tl}$ and the conjugate flux $(\hbar/e)\hat{\phi}_{nl}$ are mapped onto the 
operators\cite{note3} $\hat{Q}_l(x)$ and $\hat{\phi}_l(x)$ which are precisely described by the diagonalized Hamiltonian
\begin{eqnarray}
H_l^{bath} = \int^1_0 dx \bigg\{
{\hat{Q}_l^2(x)\over 2C_{tl}} + {\hbar^2 \over e^2} {2\over L_{tl}} 
\sin^2\bigg({\pi x \over 2}\bigg) \hat{\phi}_l^2(x) \bigg \}.
\label{Hnoise}
\end{eqnarray}
The charge (fluctuation) operator $\hat{Q}_l(x)$ and the 
phase operator $\hat{\phi}_l(x)$ obey the commutation relation
$[\hat{\phi}_l(x), \hat{Q}_l(y)/e] = i \delta(x-y)$. 
The Hamiltonian containing the couplings with the dots reads
\begin{equation}
H_{D}^{bath} = \sum_{l\sigma}\frac{e}{C_{l}}\hat{Q}_{0l} d_{l\sigma}^{\dagger} d_{l\sigma}=\sum_{l\sigma}e\delta V_{l} d_{l\sigma}^{\dagger} d_{l\sigma}.
\end{equation}
This term may arise from the extra capacitive 
coupling between each dot and the voltage
fluctuations (the quantum noise) $\delta V_{l}(t)=\hat{Q}_{0l}/C_{l}$ with $\hat{Q}_{0l}$ 
denoting the charge fluctuation operator on the given capacitor $C_{l}$, emerging from the finite
impedance $Z_l(\omega)$\cite{Markus,karyn}. According to Ref.~\onlinecite{Markus}, one can thoroughly identify $\hat{Q}_{0l} = \sqrt{2} \int^1_0 dx \cos (\pi x/2) \hat{Q}_l(x)$. 
At low frequency $\omega<\omega_{cl}=1/(R_lC_{tl})$ 
where $R_l=\sqrt{L_{tl}/C_{tl}}$, the transmission line provides an 
impedance $Z_l(\omega)=R_l/(1+i\omega/\omega_{cl})\approx R_l$. 

Below, we will absorb the $H_{D}^{bath}$ coupling into the tunneling terms through the unitary transformation\cite{Nazarov}
\begin{equation}
U=\exp\left[i\sum_{l\sigma} \delta\phi_l(t) d^{\dagger}_{l\sigma} d_{l\sigma} \right],
\end{equation}
where we have defined $\delta\phi_l(t)=(e/\hbar) \int^t dt' \delta V_{l}(t')=\sqrt{2} \int^1_0 dx \cos (\pi x/2) \hat{\phi}_l(x)$. We get $Ud_{l\sigma}U^{\dagger} = e^{i\delta\phi_l(t)} d_{l\sigma}$ and $U d_{l\sigma}^{\dagger}U^{\dagger} = e^{-i\delta\phi_l(t)} d_{l\sigma}^{\dagger}$. Moreover, exploiting
the correspondance $H' = UHU^{\dagger}+i\hbar \frac{dU}{dt}U^{\dagger}$ we realize that the couplings of the dots to the electrical baths can be completely absorbed in a redefinition of the tunneling Hamiltonian as $H'_T=H'_{SD}+H'_{DL}+{H'^*_{DL}}$ where
\begin{eqnarray}
H'_{SD} = \sum_{l\sigma} T_{SD} d^{\dagger}_{l\sigma}e^{-i\delta\phi_l(t)} \psi_{\sigma}({\bf r}_l)+h.c.= \sum_{l} H'_{SD_l}
\\ \nonumber
H'_{DL}  = \sum_{l{\bf k}\sigma} T_{DL} a^{\dagger}_{l {\bf k} \sigma} d_{l \sigma} e^{i\delta\phi_l(t)} 
= \sum_{l} H'_{D_l L_l}.
\end{eqnarray}
At a very general level, the effect of the environments can be embodied by a fluctuating phase bound to the dot's electron creation and annihilation operators such that the total Hamiltonian turns into $H'=H_0+H_T'$.

\subsection{T-matrix and general current formula}

In the quantum (zero-temperature) regime the current of two electrons passing from the superconductor via virtual dot states to the leads is formally given by
\begin{eqnarray}
\label{cur}
{I} &=& \frac{2e}{\hbar}\sum_{p,q} \rho_i W_{fi} \\ \nonumber
 &=& \frac{2e}{\hbar} \sum_{p,q} 2\pi \rho_i |\langle f_B | \langle f| T(\epsilon_i+E_B^i) |i\rangle |i_B\rangle |^2 \\ \nonumber
 &\times& \delta(\epsilon_f-\epsilon_i -E_{B}^i +E_B^f) \\ \nonumber
&=& \frac{2e}{\hbar}\sum_{p,q} 2\pi \rho_i {W}_{i,DD}{W}_{DD,f},
\end{eqnarray}
where $W_{fi}$ embodies the transition rate from the superconductor to the leads taking into account
transitions into the electrical environments; $\langle f_B|$ denotes the final (excited) state of the baths
when the injected electrons arrive in the leads, $(E_B^f-E_B^i)$ represents the energy supplied to
the environments during the EPR transportation process, ${W}_{DD,f}$ and  ${W}_{i,DD}$ stand for the transition rates from the dots to the leads and from the superconductor to the dots respectively in the presence of the fluctuations in the gate voltages, and
$\rho_i$ is the stationary occupation probability for the entire system to be in the initial ground state $|i\rangle |i_B\rangle$ where as introduced in the introduction $|i\rangle=|0\rangle_S |0\rangle_D|\mu_l\rangle$ and $|i_B\rangle$ depicts the initial state for the environments. 
Along the lines of Ref. \onlinecite{Recher}, to calculate the transition rates ${W}_{DD,f}$ and  ${W}_{i,DD}$ we resort to the T-matrix approach. The on-shell transmission or T-matrix at the energy $E$ is precisely defined as
\begin{equation}
T(E)=H_T' G(E)\frac{1}{G_0(E)}=H_T' \frac{1}{E+i\eta-H'} (E-H_0);
\end{equation}
we have introduced the Lipmann-Schwinger operators $G(E)=1/(E-H')$ and $G_0=1/(E-H_0)$ as well as the small positive real number $\eta$ that we take to zero at the end of the calculation. We rewrite the T-matrix as 
\begin{equation}
T(E)=H_T'\sum_{n=0}^{\infty} \left(\frac{1}{E-H_0+i\eta}H_T'\right)^n.
\end{equation}
It is appropriate to decompose $T(E=\epsilon_i+E_B^i)$ into the partial
T matrices ${\cal T}'$ and ${\cal T''}$.  
When two spin-entangled electrons from the superconductor leave to the dots
\begin{eqnarray}
{\cal T}'' &=&\frac{1}{i\eta+E-H_0}H'_{SD_1}\frac{1}{i\eta+E-H_0}H'_{SD_2} \\ \nonumber
&+& (1\leftrightarrow 2) \\ \nonumber
&=& \frac{1}{i\eta+E-H_0}H_{SD_1}e^{-i\delta\phi_1}\frac{1}{i\eta+E-H_0}H_{SD_2} e^{-i\delta\phi_2}\\ \nonumber
&+& (1\leftrightarrow 2),
\end{eqnarray}
where $(1\leftrightarrow 2)$ refers to the same term but exchanging the roles of the labels $1$ and $2$. ${\cal T}''$ refers to the (dissipative) crossed Andreev process. For the resonant dot $\leftrightarrow$ lead tunneling we must keep all the terms of the series
\begin{eqnarray}
\label{T'}
{\cal T}' &=& {H'_{D_1 L_1}}
\\ \nonumber
&\times& \sum_{n=0}^{\infty}
 \left(\frac{1}{E-H_0+i\eta} H'^*_{D_1 L_1} \frac{1}{E-H_0+i\eta} H'_{D_1 L_1} \right)^n \\ \nonumber
 &\times& \frac{1}{E-H_0+i\eta} H'_{D_2L_2}  \hskip 0.3cm
 \\ \nonumber
&\times& \sum_{m=0}^{\infty}  
\left(\frac{1}{i\eta+E-H_0} H_{DL}'^* \frac{1}{i\eta+E-H_0} H'_{DL}\right)^m\\ \nonumber
&+& H'_{D_2 L_2}   \sum_{n=0}^{\infty} 
 \left(\frac{1}{E-H_0+i\eta} H'^*_{D_2 L_2} \frac{1}{E-H_0+i\eta} H'_{D_2 L_2} \right)^n \\ \nonumber
&\times& \frac{1}{E-H_0+i\eta} H'_{D_1 L_1} 
\\ \nonumber
&\times&\sum_{m=0}^{\infty}  
\left(\frac{1}{E-i\eta-H_0} H_{DL}'^* \frac{1}{E-i\eta-H_0} H'_{DL}\right)^m.
\end{eqnarray}
\begin{figure}[ht]
\begin{picture}(250,100)
\leavevmode\centering\includegraphics{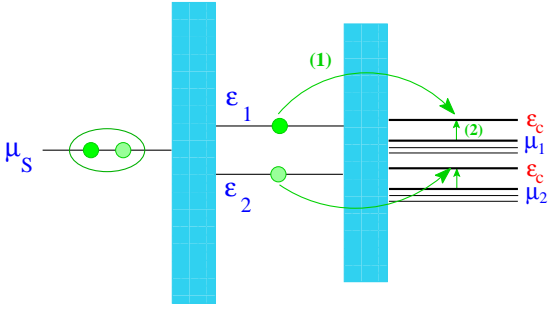}
 
 \end{picture}
\caption{(Color online) Energy conservation in the absence of noise: $2\mu_S=\epsilon_1+\epsilon_2=
 \epsilon_{\bf p} +\epsilon_{\bf q}$. When an electron on the dot 1 tunnels onto the lead 1 (1) this induces a shift of the Fermi energy $\mu_l\rightarrow \epsilon_c$ (2);  $\epsilon_c$ depicts the edge of the conduction band. In the text: $\epsilon_1\sim \epsilon_2\sim\mu_S$, $\mu_1\sim \mu_2$, and $\epsilon_c\sim \epsilon_1\sim \epsilon_2$.} 
\label{conserv}
\end{figure}

To compute the resonant dot $\leftrightarrow$ lead tunneling, we have explicitly taken into account virtual $|DD\rangle \rightarrow |DD\rangle$ transitions via the sequences $|DD\rangle \rightarrow |pD\rangle\rightarrow ... \rightarrow |DD\rangle$ or $|DD\rangle \rightarrow |Dq\rangle\rightarrow ... \rightarrow |DD\rangle$. Again the baths are insensitive to those virtual transitions which take place in a very short time and let the state of the electron system unchanged.
Here, $|pD\rangle$ stands for $a^{\dagger}_{1{\bf p}\sigma}d^{\dagger}_{2-\sigma}|i\rangle$ and implies
that the electron 1 is in lead 1 (L1) whereas the other electron resides on dot 2. We have also introduced the state $|Dq\rangle = d^{\dagger}_{1\sigma}a^{\dagger}_{1{\bf q}-\sigma}|i\rangle$ representing the state with one electron on dot 1 and the other one in lead 2 (L2). 
At this point we emphasize that virtual states with both electrons in the leads leading to $|DD\rangle \rightarrow |pD\rangle \rightarrow |pq\rangle \rightarrow |Dq\rangle \rightarrow |DD\rangle$ are suppressed by a factor $\gamma_l/\mu<1$ compared to that with only one electron in the leads and therefore can be neglected\cite{Recher}.  Recall that $|pq\rangle$ stands for $a^{\dagger}_{1{\bf p}\sigma} a^{\dagger}_{2{\bf q}-\sigma}|i\rangle$, where ${\bf p}$ is the momentum from lead 1 and ${\bf q}$ from lead 2; this is the final state $|f\rangle$ of the electrons on the leads with energy $\epsilon_f=\epsilon_{\bf q}+\epsilon_{\bf p}$. 

\subsection{``1-photon'' approximation}

In Eq. (13), we have assumed that the tunneling events are almost instantaneous implying that only one  ``photon''  is emitted in each bath during the two-particle Breit-Wigner resonance between dots and leads. In all the products $H'^*_{D_l L_l}(t)H'_{D_l L_l}(0)$ appearing in Eq. (13) we have identified
$\langle e^{-i\delta\phi_l(t)} e^{i\delta\phi_l(0)}\rangle \propto (t\omega_{cl})^{-2R_l/R_K} \rightarrow 1$,
that is ensured when the time $t$ for an electron on dot $l$ to virtually jump in lead $l$ and then to go back to dot $l$ is much shorter than $1/\omega_{cl} \sim R_l C_{tl}$. Assuming that $C_{tl}$ is large enough this should be well satisfied even for weak resistances $R_l$. Note that configurations with emission of multiple ``photons'' in the same bath would only result in small corrections in the final current. Keeping only the ``1-photon'' contribution, ${\cal T}'$ may be  summarized as\cite{noteC}
\begin{equation}
{\cal T}' = e^{i(\delta\phi_1+\delta\phi_2)} T',
\end{equation}
where $T'$ yields the same form as ${\cal T'}$ if one replaces $H'_{D_l L_l}$ by $H_{D_l L_l}$, {\it i.e.}, without dissipation $(R_1=R_2=0)$; this part is at the origin of the two-particle Breit-Wigner resonance between the dots and the leads\cite{Recher}. Since $|T_{SD}|<|T_{DL}|$ we may always rewrite ${\cal T}'' = e^{-i(\delta\phi_1+\delta\phi_2)} T''$.

Below, we will distinguish between the case of a single dissipative bath implying $R_2=0$ and therefore $\delta\phi_2=0$ and that of two (independent) baths. 
When $R_2=0$, the transition rates ${W}_{DD,f}$ and  ${W}_{i,DD}$ are given by 
\begin{eqnarray}
{W}_{DD,f} &=& \left|\langle pq| T' (\epsilon_i)|DD\rangle \langle f_B|e^{i\delta\phi_1}|D_B\rangle\right|^2 \\ \nonumber
&\times& \delta(\epsilon_f-\epsilon_1-\epsilon_2+E_B^f-E_B^D),\\ \nonumber
{W}_{i,DD} &=& \left|\langle DD|T''(\epsilon_i)|i\rangle \langle D_B|e^{-i\delta\phi_1}|i_B\rangle
\right|^2 \\ \nonumber
&\times& \delta(\epsilon_1+\epsilon_2+E_{B}^D-E_B^i-\epsilon_i).
\end{eqnarray}
Note that we have replaced $\langle pq|T'(E)|DD\rangle$ and $\langle DD|T''(E)|i\rangle$ by $\langle pq|T'(\epsilon_i)|DD\rangle$ and $\langle DD|T''(\epsilon_i)|i\rangle$ respectively; this will be thoroughly justified in Sec. II B. Furthermore, we have introduced the energy of the intermediate state of the bath $E_B^D$. The energy-conserving $\delta$ functions traduce the fact that for each tunneling process the energy of the full system including the bath is conserved or that the energy supplied to the bath is equal to the energy lost by the electrons.  The product of those two $\delta$ functions is equivalent to the $\delta$ function in Eq. (9) (dimension of current will be implicitly respected below).  

When $R_1=R_2=0$, we recover the formulas of Ref. \onlinecite{Recher}
\begin{eqnarray}
{W}_{DD,f} &=& \left|\langle pq| T' |DD\rangle\right|^2 \delta(\epsilon_f-\epsilon_1-\epsilon_2), \\ \nonumber
{W}_{i,DD} &=& \left|\langle DD|T''|i\rangle\right|^2 \delta(\epsilon_1+\epsilon_2-\epsilon_i),
\end{eqnarray}
or more precisely we obtain the following expression\cite{Recher},
\begin{eqnarray}
W_{fi}= 2\pi  \left|\langle pq| T' |DD\rangle\right|^2 \left|\langle DD|T''|i\rangle\right|^2
\delta(\epsilon_f-\epsilon_i).
\end{eqnarray}
The chemical potentials $\epsilon_1$ and $\epsilon_2$ of the quantum dots can be tuned by external
gate voltages such that the coherent tunneling of two electrons into different leads is at resonance, described by a two-particle Breit-Wigner resonance peaked at $\epsilon_1+\epsilon_2=2\mu_S=\epsilon_{\bf q}+\epsilon_{\bf p}$ (Fig. {\ref{conserv}); the situation we consider is when the two dots are close to the resonance condition $\epsilon_1=\epsilon_2=0$. We will choose energies such that $\epsilon_i=2\mu_S=0$. It is then an interesting question to understand how the quantum dissipation affects the elastic tunneling of those EPR pairs.
In the case of two independent baths, we straightforwardly generalize
\begin{eqnarray}
{W}_{i,DD} =  \left|\langle DD|T''|i\rangle\right|^2 \left|\langle D_{B_1}|e^{-i\delta\phi_1}|i_{B_1}\rangle\right|^2 \times \hskip 1cm &&\\ \nonumber
\left|\langle D_{B_2}|e^{-i\delta\phi_2}|i_{B_2}\rangle\right|^2  \delta(\epsilon_1+\epsilon_2+\sum_l E_{B_l}^D-\sum_lE_{B_l}^i-\epsilon_i), &&
\end{eqnarray}
and similarly
\begin{eqnarray}
{W}_{DD,f} =  \left|\langle pq| T' |DD\rangle\right|^2 \left|\langle f_{B_1}| e^{i\delta\phi_1} |D_{B_1}\rangle\right|^2 \times \hskip 1cm &&\\ \nonumber
\left|\langle f_{B_2}|e^{i\delta\phi_2}|D_{B_2}\rangle\right|^2 
\times \delta(\epsilon_f-\epsilon_1-\epsilon_2+\sum_l E_{B_l}^f-\sum_l E_{B_l}^D). &&
\end{eqnarray}
 We have decomposed $|D_B\rangle = |D_{B_1}\rangle |D_{B_2}\rangle$,... . 

Before to pursue, we shall discuss what is the value of $\rho_i$ from Eq. (\ref{cur}) in the presence of the environments. Similar to the noiseless case, we estimate $\rho_i=1-{\cal O}(\gamma)\rightarrow 1$ where $\gamma=\gamma_1+\gamma_2$. More precisely, the initial (gound) state $|i\rangle=|0\rangle_S |0\rangle_D|\mu_l\rangle$ is such that the highest level of the dots is unoccupied, 
there is no quasiparticle on the superconductor which is immediately fulfilled when $k_B T\ll \Delta$, and the Fermi level of the leads remain fixed to $\mu_l$. Since we consider asymmetric barriers
$|T_{DL}|>|T_{SD}|$ the most prominent transfer of electrons is between the leads and the dots. Therefore, the probability for the system to remain in the state $|i\rangle|i_{B_1}\rangle|i_{B_2}\rangle$ after a time $t$ can be estimated as $1-{\cal O}(\gamma)$. We argue that the environments will not affect this equality because as long there is no electron on the dots the environments will unambiguously remain in their ground states $|i_{B_l}\rangle$.

\section{Dissipation on dot 1}

Let us assume that $R_2=0$.
To calculate the rate for electron tunneling from the superconductor to the dots we have to evaluate
$W_{i,DD}$ from Eq. (15). Similar to Refs. \onlinecite{Nazarov,Marquardt2} we trace out environmental states leading to
\begin{eqnarray}
\hskip -0.5cm W_{i,DD} = \left|\langle DD|T''|i\rangle\right|^2 \int_{-\infty}^{+\infty} \frac{dt}{2\pi\hbar} e^{i(\epsilon_i-\epsilon_1-\epsilon_2)t/\hbar} \\ \nonumber
 \times \langle e^{i\delta\phi_1(t)} e^{-i\delta\phi_1(0)}\rangle.
\end{eqnarray}
The brackets denote an average over the initial bath ground state $|i_B\rangle$. For later convenience we like to
introduce the abbreviation $J_1(t)=\langle [\delta\phi_1(t)-\delta\phi_1(0)]\delta\phi_1(0)\rangle$ 
as well as the Fourier transform
\begin{equation}
P_1(E)=\frac{1}{2\pi\hbar} \int_{-\infty}^{+\infty} dt \exp \left[ J_1(t) +\frac{i}{\hbar} Et\right].
\end{equation}
This already permits us to write down 
\begin{eqnarray}
W_{i,DD} = \left|\langle DD|T''|i\rangle\right|^2 P_1(\epsilon_i-\epsilon_1-\epsilon_2).
\end{eqnarray}
In a similar way, we extract\cite{note2}
\begin{eqnarray}
W_{DD,f}=\left|\langle pq| T' |DD\rangle\right|^2  P_1(\epsilon_1+\epsilon_2-\epsilon_f).
\end{eqnarray}

\subsection{General discussion on $P_1(E)$}

We may interpret $P_1(E)$ as the probability to emit the energy $E$ to the electrical circuit when
transferring an electron from the superconductor to the dot 1 or from 
\begin{figure}[ht]
\begin{picture}(250,150)
\leavevmode\centering\includegraphics{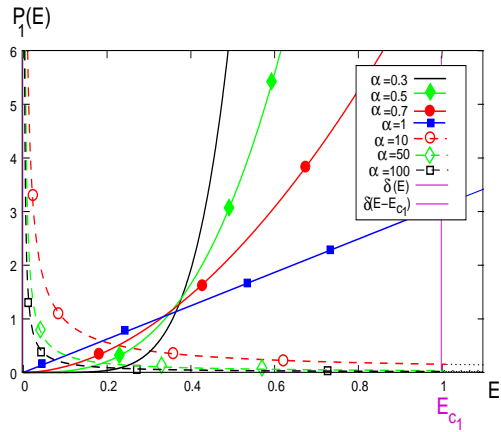}
 \end{picture}
 \vskip -0.1cm
\caption{(Color online) Function $P_1(E)$ for different values of $\alpha=\alpha_1=R_K/R_1$ from 0 to $\infty$. When $R_1$ is small one observe a blatant singularity at $E=0$ converging to a $\delta(E)$ function
 when $R_1=0$ whereas for huge $R_1$ $(\alpha_1\rightarrow 0)$ the capacitance $C_1$ plays a crucial role leading to $P_1(E)=\delta(E-E_{c1})$. We analyze the
 two limiting cases of small and very large $R_1$.} 
\label{setup}
\end{figure}
the latter to the corresponding lead 1. It is certainly useful to know more about the function $P_1(E)$ as well as $J_1(t)$. We will first envision the case of a very large resistance $R_1$. In that case the dissipative lead yields an effective impedance $Z_{eff}(\omega)=1/(R_1^{-1}+i\omega C_1)$ which tends to
$(\pi/C_1)\delta(\omega)$, $C_1$ 
being the capacitance between the dissipative lead and the dot 1. For the correlation function $J_1(t)$ this concentration of environmental modes at low frequency means that the short-time expansion\cite{Nazarov}
\begin{equation}
J_1(t)=\int_{-\infty}^{+\infty} \frac{d\omega}{\omega} \frac{\Re e Z_{eff}(\omega)}{R_K} e^{-i\omega t}= -\frac{\pi}{C_1 R_K}it,
\end{equation}
works for all times. This results in
\begin{equation}
\label{highR}
P_1(E)=\delta(E-E_{c1}),
\end{equation}
so that in order to hop onto the dot 1 an electron must transfer to the environment an amount of energy
corresponding to the charging energy $E_{c1}=e^2/(2C_{1})$ of the capacitor $C_1$; this will fatally lead to a Coulomb gap\cite{Nazarov} in the current for low  applied bias voltage $\mu\ll E_{c1}$ between the superconductor and the leads (L1 and L2). For small resistances $\alpha_1=R_K/R_1\gg 1$ 
the gate lead may be described by the frequency-independent impedance $Z_{eff}=Z_1=R_1$.
Based on the transmission line representation for the environment then we may  identify\cite{Devoret}
\begin{equation}
\label{P1}
P_1(E)=\frac{\exp(-2\gamma_e/\alpha_1)}{\Gamma(2/\alpha_1)}\frac{1}{E}\left[\frac{\pi}{\alpha_1}\frac{E}{E_{c1}}\right]^{2/\alpha_1},
\end{equation}
where $\gamma_e=0.577...$ is the Euler constant. The factor appearing may be motivated by the behavior of the correlation function $J_1(t)$ for large times\cite{Grabert}
\begin{equation}
J_1(t)=-\frac{2}{\alpha_1}\left[\ln(\alpha_1E_{c1}t/\pi\hbar)+i\frac{\pi}{2}+\gamma_e\right].
\end{equation}

The function $P_1(E)$ has been summarized through Fig. 4 with
the two distinct behaviors at low and large $R_1$.

\subsection{Discussion on tunneling matrix elements}

Now we want to properly justify the fact that the baths ``cancel out'' in the computation of the tunneling matrix elements in Eq. (15).  
 In the limit of a weak resistance $R_1$ this is straightforward since the bath 1 only absorbs a small amount of energy  during the tunneling events ($P_1(E)$ is strongly diverging at $E=0$) and therefore in all the Lipmann-Schwinger operators appearing in Eqs. (12) and (13), for a given bath state $|\alpha_B\rangle$, one can always formally replace $\langle \alpha_{B}|\sum_l H^{bath}_l |\alpha_{B}\rangle \rightarrow E_B^i$. Interestingly, we like to emphasize that for large resistances this argument still holds. In the large resistance limit, one must typically satisfy $E_B^D\rightarrow E_B^i + E_{c1}$. Thus, in order to get a finite current, one must thoroughly re-adjust the chemical potential of the SC lead such as $\epsilon_i \rightarrow \epsilon_i + E_{c1}$ with $\epsilon_i=0$; see Eq. (25). When focussing on the tunneling of a Cooper pair from the SC to the dots, hence one requires to evaluate  $\langle D_B|\langle DD| \frac{1}{E_B^i+E_{c1} - H_0 +i\eta}$. Applying $\langle D_B| H_0 = E_B^D$ we check that $\langle D_B|\langle DD| \frac{1}{E_B^i +E_{c1}- H_0 +i\eta}$ is equivalent to $\langle DD| \frac{1}{-{H}_0 +i\eta}$ by setting $H_{l}^{bath}=0$ in $H_0$. We can thus substitute $G_0$ by its expression in the absence of the bath $1$. This procedure can be extended to two baths.
For sake of clarity, calculations of $\langle pq|T'(\epsilon_i)|DD\rangle$ and $\langle DD|T''(\epsilon_i)|i\rangle$ (with the substitution $H_l^{bath}=0$ in $H_0$) are derived in Appendices B and C.

\subsection{Large resistance limit}

If we maintain the electrochemical potentials of the leads L1 and L2 and of the superconductor so that $\epsilon_1+\epsilon_2=2\mu_S=\epsilon_{\bf p} +\epsilon_{\bf q}=0$ 
due to Eq. (\ref{highR}) we
immediately infer that the current will inevitably go to zero. Indeed
\begin{eqnarray}
W_{i,DD} = \left|\langle DD|T''|i\rangle\right|^2 \delta(2\mu_S-\epsilon_1-\epsilon_2-E_{c1}) &,& \\ \nonumber
W_{DD,f} = \left|\langle pq| T' |DD\rangle\right|^2  \delta(\epsilon_1+\epsilon_2-\epsilon_f-E_{c1}) &,&
\end{eqnarray}
reflecting the dynamical Coulomb blockade phenomenon resulting from large impedances\cite{Nazarov}.
On the other hand, one could envision to symmetrically modify the electrochemical potentials of the leads L1 and L2 so that $\mu_S\rightarrow \mu_S +E_{c1}/2$ and $\mu_l\rightarrow \mu_l-E_{c1}/2$ (Fig. 5). In that case, the two $\delta$ functions above would be satisfied. Tunneling becomes possible only if the energy at disposal is equal to $E_{c1}$. 
\begin{figure}[ht]
\begin{picture}(245,112)
\leavevmode\centering\includegraphics{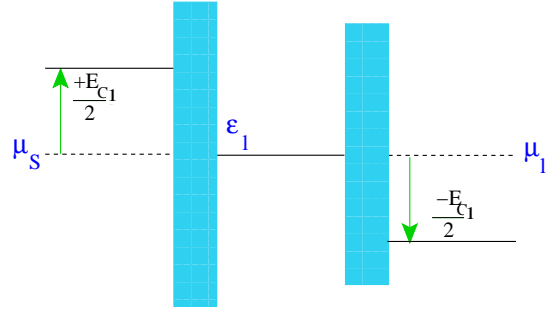}
 
 \end{picture}
\caption{(Color online) At large $R_1$, the capacitance $C_1$ must be taken into account leading to
 an additional Coulomb gap $\sim E_{c1}$; to compensate for this gap, one must thoroughly
 adjust $\mu_S\rightarrow \mu_S+E_{c1}/2$ and $\mu_l\rightarrow \mu_l-E_{c1}/2$. We will assume 
 $E_{c1}\ll \Delta$ to prevent any quasiparticle poisoning from the SC.} 
\label{setup}
\end{figure}
We assume for
this circumstance that the superconducting gap is large enough $\Delta\gg E_{c1}$ so that the superconductor is not subject to quasiparticle poisoning. Using Appendix B and mostly Eq. (\ref{T}), we value
\begin{eqnarray}
\label{val}
\hskip -0.5cm |\langle pq| T' |DD\rangle|^2 \approx |T_{DL}|^4 \left(\epsilon_1+\epsilon_2-i\eta\right)^2 \frac{16}{E_{c1}^2}{\frac{\pi \delta(\epsilon_{\bf p} +E_{c1}/2)}{\gamma_1}}. \hskip 0.3cm
\end{eqnarray}
We have used the energy conservation $\epsilon_{\bf p} +\epsilon_{\bf q} = -E_{c1}$ together with $\epsilon_1+\epsilon_2=0$ as well as $\epsilon_{\bf p}+{E_{c1}}/2 \sim \gamma_l$.
Resorting to Appendix C, we obtain the following current
\begin{equation}
I[{\delta\mu\sim E_{c1}}] = \frac{4e\gamma\gamma_S^2}{\hbar E_{c1}^2} \left(\frac{\sin({k}_F\delta {r})}{{k}_F \delta { r}}\right)^2 e^{-2\delta r/\pi \xi}.
\end{equation}
Exploiting Eq. (\ref{current}), we realize that compared to the noiseless case where $R_1=R_2=0$ the current becomes suppressed by a factor $(\gamma/E_{c1})^2$. This stems from the physical fact that shifting the electrochemical potentials of the leads such that $\mu_l\rightarrow \mu_l -E_{c1}/2$ hampers the
two-particle Breit-Wigner resonance between the dots at resonance $(\epsilon_1+\epsilon_2=0)$ and the leads. In brief, the application of a prominent
bias voltage between the dots and the leads somehow produces the ``decoherence'' of the EPR pair hence affecting the crossed Andreev current. 

\subsection{Small resistance}

We now turn our attention to the realistic situation of a small resistance so that $\alpha_1=R_K/R_1\gg 1$. Of great interest to us is to understand how the quantum noise affects the long-time coherence of the EPR pair during the two-particle Breit-Wigner process involving the dots and the leads.
An explicit calculation has been performed in Appendix D and we find an EPR current of the form
\begin{eqnarray}
\label{diss}
I \sim I[R_1=0] \frac{\exp(-2\gamma_e/\alpha_1)}{\Gamma(1+2/\alpha_1)}
\left(\frac{\pi}{\alpha_1}\right)^{\frac{2}{\alpha_1}}\left(\frac{2\mu}{E_{c1}}\right)^{\frac{2}{\alpha_1}},\hskip 0.25cm
\end{eqnarray}
where the crossed Andreev current at $R_1=0$ reads\cite{Recher}
\begin{eqnarray}
\label{current}
I[R_1=0] = \frac{4e\gamma_S^2}{\hbar \gamma}\left(\frac{\sin({k}_F\delta {r})}{{k}_F \delta { r}}\right)^2 e^{-2\delta r/\pi \xi}.
\end{eqnarray}
Note that the Breit-Wigner resonance still occurs when $\epsilon_1+\epsilon_2\sim 2\mu_S\sim \epsilon_{\bf p}+\epsilon_{\bf q}$ reflecting
the fact that for weak resistances the function $P_1(E)$ diverges at $E=0$ and therefore the bath 1
only absorbs a tiny amount of energy during the EPR transportation process. The suppression factor $\left({2\mu}/{E_{c1}}\right)^{2 R_1/R_K}$ traduces the orthogonality catastrophe arising in dissipative tunneling problems\cite{Nazarov,Devoret,Girvin,karyn,Recher4}. Now, let us return to the discussion on the decoherence (decay) of a noisy EPR Cooper pair that is described, after averaging over the bath degrees of freedom, by the dot state $\langle e^{-i\delta\phi_1} \rangle [d^{\dagger}_{1\uparrow}d^{\dagger}_{2\downarrow}-d^{\dagger}_{1\downarrow}d_{2\uparrow}^{\dagger}]|i\rangle$. In the weak resistance realm\cite{Nazarov}, we can evaluate
\begin{equation}
\langle e^{-i\delta\phi_1(t)} \rangle = \exp\left[{-\frac{1}{\alpha_1} \ln\left(\frac{E_{c1}\alpha_1 t}{\pi\hbar}\right)}\right].
\end{equation}
At very long-times $t\gg \hbar \pi/(\alpha_1 E_{c1})$ or very low energies $\mu\ll E_{c1}$, by coupling to the environment the noisy EPR pair looses its phase coherence affecting drastically the efficiency of the two-particle Breit-Wigner resonance between the dots and the leads L1 and L2. It is worth to note the similitude with the power-law suppression in the Andreev entangler with Luttinger leads\cite{Recher2,Smitha}. For larger resistances, at low $\mu$, we observe a precursor effect of the dynamical Coulomb gap mentioned above.

\section{Two independent baths}

We can straightforwardly generalize the previous analysis to the case of two (independent) environments assuming $R_2 \neq 0$. The two-particle Breit-Wigner transport between the leads and the dots gets modified as
\begin{eqnarray}
W_{DD,f} = \left|\langle pq|T'|DD\rangle\right|^2 P_{12}(\epsilon_1+\epsilon_2-\epsilon_{\bf p}-\epsilon_{\bf q}),
\end{eqnarray}
where we have introduced
\begin{eqnarray}
P_{12}(E)=\int_{-\infty}^{+\infty} \frac{dt}{2\pi\hbar} e^{i E t/\hbar} e^{J_1(t)+J_2(t)},
\end{eqnarray}
and $J_2(t)=\langle [\delta\phi_2(t)-\delta\phi_2(0)]\delta\phi_2(0)\rangle$. When the two resistances are larger than $R_K$, $P_{12}(E)=\delta(E-E_{c1}-E_{c2})$ where $E_{c2}=e^2/(2C_2)\sim E_{c1}$ is the charging energy of the capacitor $C_2$. To get a finite current between the superconductor and the leads 1 and 2 one needs to re-adjust $2\mu_S\rightarrow 2\mu_S + E_{c1} + E_{c2}$, $\mu_1 \rightarrow \mu_1 - E_{c1}$, and $\mu_2 \rightarrow \mu_2 - E_{c2}$ (we assume $E_{c1}\sim E_{c2}$ so that L1 and L2 are kept at the same electrochemical potential). We find
\begin{equation}
I[{\delta\mu\sim 2E_{c1}}] = \frac{I[R_1=0]}{4}\gamma^2\left(\frac{1}{E_{c1}}+\frac{1}{E_{c2}}\right)^2.
\end{equation}
We can observe a huge suppression factor $\sim(\gamma/E_{c1} +\gamma/E_{c2})^2$.
When the two resistances are much smaller than $R_K$, which is the situation of most interest, we get
\begin{equation}
P_{12}(E)= \frac{\exp(-2\gamma_e/\alpha)}{\Gamma(2/\alpha)}\frac{1}{E}\left[\frac{\pi}{\alpha}\frac{E}{E_{c1}}\right]^{2/\alpha},
\end{equation}
where
\begin{equation}
\alpha^{-1}=\alpha_1^{-1}+\alpha_2^{-1}=\frac{R_1}{R_K} + \frac{R_2}{R_K}.
\end{equation}
We obtain a result identical to that of a unique weakly-resistive bath with $\alpha_1\rightarrow \alpha$
(see Appendix D). Recall that the orthogonality catastrophe becomes more pronounced and the suppression factor in the crossed Andreev current now follows $(2\mu/E_{c1})^{2/\alpha}= (2\mu/E_{c1})^{2/\alpha_1 +2/\alpha_2}$; consult Eq. (D15). Finally one could envision to investigate the asymmetric case where one resistance is well prominent, {\it e.g.}, $R_1$, and the other is weak but nonzero resulting in
\begin{equation}
P_{12}(E)= \frac{\exp(-2\gamma_e/\alpha_2)}{\Gamma(2/\alpha_2)}\frac{1}{E}\left[\frac{\pi}{\alpha_2}\frac{E-E_{c1}}{E_{c1}}\right]^{2/\alpha_2}. 
\end{equation}
It is important to visualize that in that case $P_{12}(E)$ yields a visible singularity at $E=E_{c1}$ (which is reminiscent of the situation where $R_1$ is large and $R_2=0$) and
therefore to get a current through the structure again one must re-adjust $\mu_S\rightarrow \mu_S+E_{c1}/2$ and $\mu_l\rightarrow \mu_l-E_{c1}/2$. We replace $\left|\langle pq|T'|DD\rangle\right|^2$  by its value in Eq. (\ref{val}) leading to
\begin{equation}
 I = I[\delta\mu\sim E_{c1}] \int_{\mu_l}^{\epsilon_c} d\epsilon_{\bf q} P_2(\epsilon_1+\epsilon_2 - \mu_l-\epsilon_{\bf q}),
\end{equation}
and therefore
\begin{eqnarray}
\hskip -0.5cm I[\delta\mu\sim E_{c1},\alpha_2\gg 1] \sim {I[\delta\mu\sim E_{c1}]} \frac{\exp(-2\gamma_e/\alpha_2)}{\Gamma(1+2/\alpha_2)}
\left(\frac{\mu}{E_{c1}}\right)^{\frac{2}{\alpha_2}}. \hskip 0.2cm
\end{eqnarray}
The crossed Andreev current becomes markedly suppressed by a factor proportional to $(\gamma/E_{c1})^2 (2\mu/E_{c1})^{2/\alpha_2}$ and $\mu\ll E_{c1}$ is the applied bias voltage before the re-adjustment
$\mu_S\rightarrow \mu_S+E_{c1}/2$ and $\mu_l\rightarrow \mu_l-E_{c1}/2$.

\section{Parasitic direct Andreev processes}

Thus far, we have completely omitted processes allowing the two electrons forming 
the Cooper pair to jump onto the same dot. In absence of thermal effects, the latter are somehow reduced due to the traditional Coulomb blockade phenomenon on the dots as well as the superconducting gap. Below, we will precisely discuss the quantum noise effect on those processes as well as the efficiency condition(s) of the noisy Andreev entangler.

In the absence of quantum noise, direct Andreev processes (where the two electrons take the same
dot) are suppressed by a factor $(\gamma_l/U)^2$ and/or $(\gamma_l/\Delta)^2$ compared to the crossed Andreev process as a consequence of the 
Coulomb blockade and the superconducting gap\cite{Recher}. Nevertheless, those direct Andreev processes do not suffer from a suppression resulting from the spatial separation of the quantum dots. We will use the terminology of Recher {\it et al.}\cite{Recher} by identifying two distinct direct Andreev processes: (I) In the first step, one electron tunnels from the
superconductor to, say, dot 2, and in a second step the second electron also tunnels to dot 2. There are now two electrons on dot 2 which costs the Coulomb repulsion energy $U$; this virtual state is suppressed by $1/U$. Hence the two electrons leave dot 2 and tunnel to lead 2 (L2) one after the other. (II) There is a competing process that avoids double occupancy on dots but leaves an excitation on the superconductor that costs $1/\Delta$. Here, one electron tunnels to, say, dot 2 and then the same electron tunnels further into lead 2. Finally, the second electron tunnels from the superconductor via dot 2 into lead 2.

\begin{figure}[ht]
\begin{picture}(250,230)
\leavevmode\centering\includegraphics{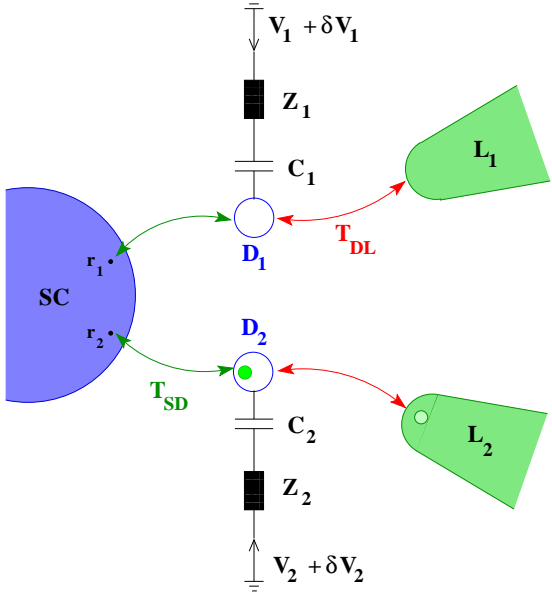}
 
 \end{picture}
 \vskip 0.1cm
\caption{(Color online) Due to the Coulomb blockade direct Andreev parasitic processes are (diminished 
 and) triggered by single-particle resonances between dot $\leftrightarrow$ lead. Noise will also affect the low-energy coherence of the single-particle current.} 
\label{setup}
\end{figure}

We first concentrate on the tunneling process (II) including the quantum noise. The current $I^{(II)}$
from the superconductor to the final lead state takes the form
\begin{equation}
I^{(II)} = \frac{2e}{\hbar} \sum_{{\bf p}, {\bf p}'} \sum_l 2\pi W_{i,Dp''}^{(II)} W_{Dp'',f}^{(II)}.
\end{equation}
We define the transition rate from $|D p''\rangle$ to $|f\rangle$  as
$W_{Dp'',f}^{(II)} = |w_{D p'', f}^{(II)} | ^2 \delta(\epsilon_f+E_B^f - \epsilon_l - \epsilon_{{\bf p}''} - E_B^{D p''})$ and
\begin{eqnarray}
 w_{Dp'',f}^{(II)} = \sum_{{\bf p}'' \sigma} \langle f_B|\langle f | H_{DL}' | D p'' \sigma \rangle |D_B p'' \rangle 
\langle D_B p'' | \langle D p'' \sigma |  && \hskip 0.2cm
\\ \nonumber 
\times \sum_{n=0}^{\infty}\left(\frac{1}{i\eta - H_0} H'^*_{DL} \frac{1}{i\eta - H_0} H'_{DL}\right)^n | D p'' \sigma \rangle | D_B p'' \rangle. &&
\end{eqnarray}
The final state $|f\rangle$ with two electrons in the same lead in the singlet state obeys $|f\rangle
= (1/\sqrt{2}) (a^{\dagger}_{{\bf p}\uparrow}a^{\dagger}_{{\bf p}' \downarrow} -  
a^{\dagger}_{{\bf p}\downarrow}a^{\dagger}_{{\bf p}' \uparrow}) |i\rangle$ and we have introduced
the intermediate state $|Dp''\rangle = d^{\dagger}_{-\sigma} a^{\dagger}_{{\bf p}'' \sigma} |i\rangle$
with one electron on, say, dot 2, with spin $-\sigma$ and the other one already in lead 2 with spin $\sigma$ and momentum ${\bf p}''$. The process where the first electron leaves
the superconductor and tunnels to the lead 2 and where the second electron tunnels onto the dot 2 has to be accomplished in a very short-time $\sim \hbar/ \Delta$ and thus we assume that this process is {\it instantaneous} implying that the bath only reacts when the second electron resides on the dot 2. 
The intermediate state of the bath with one electron on the dot and the other in the lead has been denoted $|D_B p'' \rangle$ and its energy $E_B^{Dp''}$. We also decompose $W_{i,Dp''}^{(II)}= |w_{i,Dp''}^{(II)}|^2
\delta(\epsilon_{{\bf p}''}+\epsilon_l +E_B^{Dp''} - \epsilon_i -E_B^i)$ where
\begin{eqnarray}
w_{i,Dp''}^{(II)} = \langle D_B p'' |  \langle D p'' \sigma |  \frac{1}{i\eta-H_0} H_{SD}' \frac{1}{i\eta - H_0} H_{DL}'
\\ \nonumber
\times \frac{1}{i\eta-H_0} H_{SD}' |i\rangle |i_B\rangle.
\end{eqnarray}
In the Lipmann-Schwinger operators, again we have taken into account that the bath's energies cancel
out (in $H_0$ now one must equate $H_l^{bath} =0$). From Sec. II B, we know that this is always justified in the weak-resistance limit which will be the situation of interest below.
Note that a tunnel process from the state $|i\rangle$ to the state $|Dp''\rangle$ does not have to be resummed further since this would lead either to a double occupancy of the dot that is suppressed by $1/U$ or to a state with two electrons simultaneously in the lead that is suppressed by a factor\cite{Recher} $\gamma_l/\mu<1$. For convenience, we have suppressed the label $l=1,2$ in the Hamiltonians 
$H'_{D_l L_l}$ and $H_{SD_l}'$ above. In the presence of voltage noise, we can yet decompose
\begin{eqnarray}
\label{w}
W_{Dp'',f}^{(II)} &=& W_{D p'', f}^{(II)o} \left|\langle f_B| e^{i\delta\phi_l} |D_B {p''}\rangle\right|^2 
\\ \nonumber
&\times& \delta(\epsilon_f+E_B^f - \epsilon_l - \epsilon_{{\bf p}''} - E_B^{D p''}) \\ \nonumber
&=& W_{Dp'',f}^{(II)o} P_l(\epsilon_{{\bf p}''} +\epsilon_l - \epsilon_f),
\end{eqnarray}
as well as
\begin{eqnarray}
\label{w2}
W_{i,Dp''}^{(II)} &=& W_{i,Dp''}^{(II)o} \left|\langle D_B p''| e^{-i\delta\phi_l} |i_B\rangle\right|^2 
\\ \nonumber
&\times& \delta(\epsilon_{{\bf p}''}+\epsilon_l +E_B^{Dp''} - \epsilon_i -E_B^i) \\ \nonumber
&=& W_{i,Dp''}^{(II)o} P_l(\epsilon_i - \epsilon_{{\bf p}''} - \epsilon_l),
\end{eqnarray}
and $(...)^{(II) o}$ are exactly the transition rates occurring in
the absence of dissipative effects. Note that the product of Hamiltonians $H_{SD}' H'_{DL} H'_{SD}$
is equivalent to $e^{-i\delta\phi_l} H_{SD} H_{DL} H_{SD}$. Moreover we identify
\begin{eqnarray}
\label{amp}
W_{D p'', f}^{(II) o}  W_{i,Dp''}^{(II) o}  &=& - \frac{2^{3/2} \nu_S (T_{SD} T_{DL})^2 }{\Delta (\epsilon_{\bf p} + \epsilon_l -i\gamma_l/2)} \\ \nonumber
&\times& \frac{(\epsilon_{\bf p} +\epsilon_
{{\bf p}'})/2 +\epsilon_l -i\gamma_l/2}{\epsilon_{{\bf p}'} +\epsilon_l - i\gamma_l/2}.
\end{eqnarray} 
The momentum ${\bf p''}$ does not appear in the final result because one must satisfy either $\delta_{{\bf p},{\bf p}''}$ or $\delta_{{\bf p},{\bf p}'}$. In the absence of noise where $\epsilon_{\bf p} +\epsilon_{{\bf p}'}=2\mu_S=0=\epsilon_{{\bf p}''} +\epsilon_l$ we explicitly recover the formula of Ref. \onlinecite{Recher}. Now, repeating the same calculation for the process (I) we find that $W_{Dp'',f}^{(I)}=W_{Dp'',f}^{(II)}$ and furthermore $w_{i,Dp''}^{(I)}$ obeys
\begin{eqnarray}
w_{i,Dp''}^{(I)} = \langle D_B p'' |  \langle D p'' \sigma |  \frac{1}{i\eta-H_0} H_{DL}' \frac{1}{i\eta - H_0} H_{SD}'
\\ \nonumber
\times \frac{1}{i\eta-H_0} H_{SD}' |i\rangle |i_B\rangle.
\end{eqnarray}
Compared to $w_{i,Dp''}^{(II)}$ the order of tunneling events has changed.
$W_{i,Dp''}^{(I)}$ has a form similar to Eq. (\ref{w2}) and the amplitude product $W_{D p'', f}^{(I) o}  W_{i,Dp''}^{(I) o}$ is given by Eq. (\ref{amp}) but with $\Delta$ being replaced by the Coulomb gap $U/\pi$.

In Appendix E, we have summarized the calculations of the direct Andreev current $I'=I^{(I)} + I^{(II)}$ and the main result is that for relatively weak resistances 
\begin{eqnarray}
I'[\alpha_l\gg 1] = \frac{I'[R_l=0]}{2} \sum_l \frac{\exp(-2\gamma_e/\alpha_l)}{\Gamma(1+2/\alpha_l)} \left(\frac{2\mu \pi}{\alpha_l E_{cl}}\right)^{2/\alpha_l}, \hskip 0.2cm
\end{eqnarray}
and for the noiseless case we again agree with Ref. \onlinecite{Recher}
\begin{eqnarray}
I'[R_l=0] = \frac{2e \gamma_S^2 \gamma}{\hbar {\cal E}^2};
\end{eqnarray}
${\cal E}^{-1}=1/(\pi \Delta) +1/U$ has been already mentioned in the introduction. It is important
to stress that even though the direct Andreev current is suppressed by a factor $(\gamma_l/{\cal E})^2$
compared to the crossed Andreev current $I$, the former is less sensitive to voltage noise in the sense
that this is only affected by an extra factor $\sim (\mu/E_{c1})^{2/\alpha_l}$ as opposed to $(\mu/E_{c1})^{2/\alpha_1+2/\alpha_2}$ for the crossed Andreev current $I$ given explicitly in Eq. (D15) assuming two environments. This stems from the fact that the crossed Andreev current involves a {\it two-particle} Breit-Wigner resonance and thus is more sensitive to voltage noise than the direct Andreev processes which demand that one electron instantaneously leaves to the lead $l$ hence producing a {\it single-particle} Breit-Wigner type transport through, {\it e.g.}, $W_{D p'',f}^{(II)}$. {\it This is the uppermost issue of our paper}. 
 
 Note, in passing, that for the asymmetric case where $R_2=0$ strictly whereas $R_1$ would be finite (but much smaller than $R_K$), which means that dissipation only
 concerns electrons residing on dot 1, then we easily extract
 \begin{eqnarray}
\hskip -0.2cm \frac{I'[\alpha_1\gg 1]}{I'[R_l=0]} = \frac{1}{2} \left[ 1+\frac{\exp(-2\gamma_e/\alpha_1)}{\Gamma(1+2/\alpha_1)} \left(\frac{2\mu \pi}{\alpha_1 E_{c1}}\right)^{2/\alpha_1}\right].
 \end{eqnarray}
When the two electrons forming the injected Cooper pair take the dot 2, the direct Andreev current is 
equivalent to that of the noiseless case.  Moreover the direct Andreev current stemming from the passage of the two spin-entangled electrons through dot 1 is affected by the noise in the same manner as the crossed Andreev current.

\section{Conclusion}

In brief, in the setup of Fig. 1 we have thoroughly investigated the effect of voltage noise produced by the electrical circuits in the vicinity of the quantum dots\cite{Markus,karyn} on the transportation of nonlocal charged-2e Cooper pairs (spin-based EPR pairs). 
We emphasize that even though electron spins  in a semiconductor environment show unusually long dephasing times approaching microseconds and can be transported phase-coherently over distances exceeding $100\mu m$\cite{Asc}, in the realistic dot-based Andreev entangler introduced in Ref. \onlinecite{Recher}
the voltage noise may affect the transportation of those EPR pairs through the charge degrees of freedom. Although the spin entanglement is preserved at long times, as a result of the entanglement of the charge 2e with the electromagnetic noise the (noisy) Cooper pair object inevitably decays at long times. Assuming almost instantaneous tunneling events, we have been able to build
a ``P(E) theory'' for this problem along the lines of Ref. \onlinecite{Marquardt2}. More precisely, when investigating
the Breit-Wigner resonances between the dots and the leads, we have kept only
the dominant  one ``photon'' contribution from each bath. 
For moderate and symmetric resistances $R_1\sim R_2 \ll R_K$, the condition for the noisy EPR entangler to be efficient ($I/I' >1$) reads: 
\begin{equation}
\left(\frac{\cal E}{\gamma}\right) \left(\frac{\mu}{E_{c1}}\right)^{R_l/R_K} > (k_F \delta r).
\end{equation}
Here, we have considered that $\delta r<\pi \xi$ and two {\it independent} baths. 
It is important to recall that the crossed Andreev current is triggered by an ``EPR-pair'' Breit-Wigner resonance between the two leads and the two dots and is therefore (slightly) more sensitive to voltage noise than the parasitic direct Andreev processes which only involve a single-particle Breit-Wigner resonance between, say, dot $1$ and lead $1$ (one electron has been instantaneously transmitted to the lead, {\it e.g.}, due to the Coulomb blockade effect). 
Assuming that the Coulomb gap is large enough to satisfy this renormalized efficiency
condition, it would be interesting to probe experimentally the long-time decoherence of the EPR pair through the orthogonality catastrophe factor $(2\mu/E_{c1})^{4 R_l/R_K}$ appearing in the crossed Andreev current. We like to emphasize that one could envision to exploit the dissipative GaAs heterostructures of Ref. \onlinecite{Rimberg} to build a quantum dot in the proximity of a two-dimensional electron gas in low density which then serves as a {\it tunable} source of dissipation; interestingly, the resistance of the envionment can reach few $k\Omega$. At this step, it is certainly important to also give our opinion on the case where the two dots would be subject to the voltage fluctuations of the {\it same} environment possessing a resistance $R$. We find that Eq. (19) would turn into $W_{DD,f} = |\langle pq| T'|DD\rangle|^2 |\langle f_B| e^{2i\delta\phi} |D_B\rangle|^2 \delta(\epsilon_f-\epsilon_1-\epsilon_2 +E_B^f-E_B^D)$. It follows that the crossed Andreev current would be subject to a more dramatic suppression $\sim (2\mu/E_{c1})^{8 R/R_K}$ whereas the direct Andreev current would exhibit the same power-law suppression as the two independent bath case. The efficiency condition of the entangler turns into
\begin{equation}
\left(\frac{\cal E}{\gamma}\right)^2 \left(\frac{\mu}{E_{c1}}\right)^{6R/R_K} > (k_F \delta r)^2.
\end{equation}

It is relevant to note that the setup of interest to us is quite different from that of a BCS-superconductor directly coupled to two highly-resistive normal leads being described by two electromagnetic environments\cite{Recher4}. In that case, similar to Luttinger liquid leads\cite{Recher3,Smitha}, tunneling of two spin-entangled electrons into the same lead is diminished compared to the crossed Andreev process where the pair splits and each electron tunnels into different leads. The reason is that when a charge $2e$ tunnels into the same lead $l$ the tunneling process is accompanied by a phase $e^{-i2\phi_l}$; this doubling of the phase leads to a more prominent suppression of current compared to the case where the two electrons take different leads. We insist on the fact that in the setup based on quantum dots, when the two electrons tunnel into the same lead the current cannot be triggered by charges $2e$ due to the Coulomb blockade; the main process between the dot $l$ and the lead $l$ is a single-particle Breit-Wigner resonance. 

Note in passing that similar conclusions would typically arise when including the noise in the normal leads L1 and L2 in the setup of Fig. 1; more precisely, in that
case we should carefully replace $a_{l{\bf k}\sigma}\rightarrow a_{l{\bf k}\sigma}e^{i\delta\phi_l(t)}$ in Eq. (8) and hence the same conclusions could be
derived.

Since the spin entanglement is not really affected by the electrical noise, one
could ask whether it would be possible to detect the spin entanglement of a noisy EPR pair despite the suppressed transmission probability of the latter at low voltage or long time. For example, Ref. \onlinecite{Samuelsson} envisions to introduce a beam-splitter and focus on zero-frequency current correlations. Those quantities will be affected in a similar way as the currents ({\it e.g.}, the total noise of the current flowing out of the superconductor is related to currents
through the Schottky's result\cite{Samuelsson}) and thus the detection of current correlations becomes highly dependent on the bias voltage. However, since the currents and the zero-frequency current correlations are affected in a similar way, we must admit that the Fano factors given in \onlinecite{Samuelsson} should not be modified that might give some hope to detect the nonlocal spin entanglement.

Finally, the question whether the EPR pair can survive if the dots are pushed away from resonance and are singly-occupied, is an interesting question that would be worthwhile to investigate further. The (related) structure with a double dot in the Coulomb blockade regime coupled to two superconducting leads\cite{Choi} is well known to induce an antiferromagnetic coupling between the dots.

{\it Acknowledgments:} The authors are very grateful to M. B\"{u}ttiker, D. Feinberg, Mei-Rong Li,  D. Loss, and E. Sukhorukov, for useful comments on the paper. K.L.H. is also grateful to P. Recher for a very careful reading of the manuscript as well as very constructive discussions. K.L.H. was supported by CIAR, FQRNT, and NSERC.

\appendix

\section{BCS notations}

The s-wave superconductor is described by the BCS theory. The BCS Hamiltonian takes the form
\begin{equation}
H_S=\sum_{{\bf k}\sigma} E_{\bf k} \gamma^{\dagger}_{{\bf k}\sigma} \gamma_{{\bf k}\sigma},
\end{equation}
$E_{\bf k}=\sqrt{\xi_{\bf k}^2+\Delta^2}$ being the quasiparticle energy and $\Delta$ the superconducting
gap. The quasiparticle operator $\gamma_{{\bf k}\sigma}$ is related to the electron annihilation and creation operators $c_{{\bf k}\sigma}$ and $c^{\dagger}_{{\bf k}\sigma}$ through the Bogoliubov transformation
\begin{eqnarray}
c_{{\bf k}\uparrow} = u_{\bf k} \gamma_{{\bf k}\uparrow} + v_{\bf k} \gamma^{\dagger}_{{-\bf k}\downarrow}, \\ \nonumber
c_{-{\bf k}\downarrow} = u_{\bf k} \gamma_{-{\bf k}\downarrow} - v_{\bf k} \gamma_{{\bf k}\uparrow}^{\dagger},
\end{eqnarray}
where the coherence factors $u_{\bf k}=\sqrt{1+(\xi_{\bf k}/E_{\bf k})}/\sqrt{2}$ and $v_{\bf k}=\sqrt{1-(\xi_{\bf k}/E_{\bf k})}/\sqrt{2}$ have been introduced and $\xi_{\bf k}=\epsilon_{\bf k}-\mu_S$ is the normal state single-electron 
energy counted from the Fermi level $\mu_S$. We choose energies such that $\mu_S=0$.
$\psi_{\sigma}({\bf r}_l)$ annihilates an electron in the superconductor at the site ${\bf r}_l$ and $\psi_{\sigma}({\bf r}_l)$ is related to $c_{{\bf k}\sigma}$ by the Fourier transform $\psi_{\sigma}({\bf r}_l)=\sum_{\bf k} e^{i{\bf k} {\bf r}_l} c_{{\bf k}\sigma}$.
In our calculations, we will have to compute quantities like $\langle i| \gamma_{{\bf k}\sigma} \psi_{-\sigma}({\bf r}_l)= \sum_{{\bf k}'}e^{i{\bf k}' {\bf r}_l}\langle i| \gamma_{{\bf k}\sigma} c_{{\bf k}'-\sigma}$. The only terms
which are non-zero should be proportional to  $\langle i| \gamma_{{\bf k}\sigma}\gamma_{{\bf k}\sigma}^{\dagger}=\langle i|$. Hence we infer that we shall select $c_{{\bf k}'\sigma}=c_{-{\bf k}\sigma}\rightarrow \epsilon_{\sigma} v_{\bf k} \gamma^{\dagger}_{{\bf k}\sigma}$ in the equations (A2) above resulting in $\langle i| \gamma_{{\bf k}\sigma} c_{{\bf k}'-\sigma}=\langle i| v_{\bf k} \delta({\bf k+k}')\epsilon_{\sigma}$ $(\epsilon_{\sigma}=\pm$ for $\sigma=\uparrow,\downarrow$). We evaluate $\langle i| \psi_{\sigma}({\bf r}_l) \gamma^{\dagger}_{{\bf k}\sigma}=\sum_{{\bf k}''}\langle i| e^{i{\bf k}'' {\bf r}_l}c_{{\bf k}''\sigma}\gamma^{\dagger}_{{\bf k}\sigma}$ in a similar way and we easily extract
$\langle i| c_{{\bf k}''\sigma} \gamma^{\dagger}_{{\bf k}\sigma} = \langle i| u_{\bf k} \delta({\bf k-k}'')$.

\section{Calculation of $\langle pq|T'|DD\rangle$}

Our aim is now to present a detailed calculation of  $\langle pq|T'|DD\rangle$ without imposing
$\epsilon_{\bf p}+\epsilon_{\bf q}=2\mu_S=0$ (equality stemming from the energy conservation in the absence of dissipation). First, we can rigorously simplify
\begin{eqnarray}
\langle DD | \sum_{n=0}^{\infty} \left(\frac{1}{i\eta-H_0} H^*_{DL} \frac{1}{i\eta-H_0} H_{DL}\right)^n|DD\rangle \\ \nonumber
\hskip -1cm = \frac{1}{1-\langle DD | \frac{1}{i\eta-H_0} H_{DL}^* \frac{1}{i\eta-H_0} H_{DL}|DD\rangle}.
\end{eqnarray}
Hence, we exploit 
 $\langle DD| \frac{1}{i\eta-H_0} = \langle DD| \frac{1}{i\eta-\epsilon_1-\epsilon_2}$ as well as $\langle DD|(H_{D_1L_1}^*\frac{1}{i\eta-H_0}H_{D_1L_1}+H_{D_2L_2}^*\frac{1}{i\eta-H_0}H_{D_2L_2})|DD\rangle=|T_{DL}|^2 
\sum_{l,{\bf k}}\frac{1}{i\eta-\epsilon_{\bf k}-\epsilon_l}$ resulting in
\begin{eqnarray}
\hskip -0.5cm \langle DD | \frac{1}{i\eta-H_0}H_{DL}^*\frac{1}{i\eta-H_0} H_{DL}|DD\rangle = 
 \frac{\Sigma}{i\eta-\epsilon_1-\epsilon_2},
 \end{eqnarray}
 where we have introduced the self-energy
 \begin{equation}
\Sigma=|T_{DL}|^2 \sum_{l,{\bf k}} (i\eta-\epsilon_l-\epsilon_{\bf k})^{-1}.
\end{equation}
Akin to Ref. \onlinecite{Recher}, we can straightforwardly decompose $\Sigma=\Re e\Sigma-i\gamma/2$ where $\gamma=\gamma_1+\gamma_2$ and 
$\Re e\Sigma \sim \gamma_l \ln(\epsilon_c/\mu_l)$ can be neglected assuming that the renormalization
of the energy level is small. This leads to the expression
\begin{eqnarray}
\langle DD | \sum_{n=0}^{\infty} \left(\frac{1}{i\eta-H_0} H^*_{DL} \frac{1}{i\eta-H_0} H_{DL}\right)^n|DD\rangle
\\ \nonumber
= \frac{1}{1-\frac{-i\gamma/2}{i\eta-\epsilon_1-\epsilon_2}}=\frac{\epsilon_1+\epsilon_2-i\eta}{\epsilon_1+\epsilon_2-i\gamma/2}.
\end{eqnarray}

Similar results hold for the one-particle resummation
\begin{equation}
\langle pD | \sum_{n=0}^{\infty} \left(\frac{1}{i\eta-H_0} H^*_{D_2L_2}) \frac{1}{i\eta-H_0} H_{D_2L_2}\right)^n|pD\rangle
\end{equation}
then providing
\begin{eqnarray}
\langle pD | \sum_{n=0}^{\infty} \left(\frac{1}{i\eta-H_0} H^*_{D_2L_2} \frac{1}{i\eta-H_0} H_{D_2L_2}\right)^n|pD\rangle \\ \nonumber
= \frac{1}{1-\frac{-i\gamma_2/2}{i\eta-\epsilon_{\bf p}-\epsilon_2}}=\frac{\epsilon_{\bf p}+\epsilon_2-i\eta}{\epsilon_{\bf p}+\epsilon_2-i\gamma_2/2}.
\end{eqnarray}
Again, $|pD\rangle$ stands for $a^{\dagger}_{1{\bf p}\sigma}d^{\dagger}_{2-\sigma}|i\rangle$ and implies
that the electron 1 is in lead 1 (L1) whereas the other electron resides on dot 2. We can also introduce the state $|Dq\rangle = d^{\dagger}_{1\sigma}a^{\dagger}_{2{\bf q}-\sigma}|i\rangle$ representing the state
with one electron on dot 1 and the other one in lead 2 (L2) leading to
\begin{eqnarray}
\langle Dq | \sum_{n=0}^{\infty} \left(\frac{1}{i\eta-H_0} H^*_{D_1L_1} \frac{1}{i\eta-H_0} H_{D_1L_1}\right)^n|Dq\rangle \\ \nonumber
= \frac{1}{1-\frac{-i\gamma_1/2}{i\eta-\epsilon_{\bf q}-\epsilon_1}}=\frac{\epsilon_{\bf q}+\epsilon_1-i\eta}{\epsilon_{\bf q}+\epsilon_1-i\gamma_1/2}.
\end{eqnarray}
We can now proceed and compute  
\begin{eqnarray}
& &\langle pq|T'|DD\rangle =\langle pq|H_{D_1 L_1} |Dq\rangle \\ \nonumber
&\times&
\langle Dq | \sum_{n=0}^{\infty} \left(\frac{1}{i\eta-H_0} H^*_{D_1L_1} \frac{1}{i\eta-H_0} H_{D_1L_1}\right)^n|Dq \rangle \\ \nonumber
&\times& \langle Dq|\frac{1}{i\eta-H_0}H_{D_2 L_2}|DD\rangle \\ \nonumber
&\times&
\langle DD | \sum_{m=0}^{\infty} \left(\frac{1}{i\eta-H_0} H^*_{DL} \frac{1}{i\eta-H_0} H_{DL}\right)^m|DD\rangle
\\ \nonumber
&+& \langle pq|H_{D_2 L_2} |pD\rangle \\ \nonumber
&\times&
\langle pD | \sum_{n=0}^{\infty} \left(\frac{1}{i\eta-H_0} H^*_{D_2 L_2} \frac{1}{i\eta-H_0} H_{D_2 L_2}\right)^n|pD \rangle \\ \nonumber
&\times& \langle pD|\frac{1}{i\eta-H_0}H_{D_1 L_1}|DD\rangle \\ \nonumber
&\times&
\langle DD | \sum_{m=0}^{\infty} \left(\frac{1}{i\eta-H_0} H^*_{DL} \frac{1}{i\eta-H_0} H_{DL}\right)^m|DD\rangle.
\end{eqnarray}
We obtain
\begin{eqnarray}
& &\langle pq|T'|DD\rangle =T_{DL} \frac{\epsilon_{\bf q}+\epsilon_1-i\eta}{\epsilon_{\bf q}+\epsilon_1-i\gamma_1/2} \\ \nonumber
&\times& \left(\frac{T_{DL}}{i\eta-\epsilon_1-\epsilon_{\bf q}}\right)\frac{\epsilon_1+\epsilon_2-i\eta}{\epsilon_1+\epsilon_2-i\gamma/2} \\ \nonumber
&+& T_{DL} \frac{\epsilon_{\bf p}+\epsilon_2-i\eta}{\epsilon_{\bf p}+\epsilon_2-i\gamma_2/2} 
\left(\frac{T_{DL}}{i\eta-\epsilon_2-\epsilon_{\bf p}}\right)\frac{\epsilon_1+\epsilon_2-i\eta}{\epsilon_1+\epsilon_2-i\gamma/2},
\end{eqnarray}
and therefore
\begin{eqnarray}
\hskip -0.9cm \langle pq|T'|DD\rangle = -T_{DL}^2\frac{1}{\epsilon_{\bf p}+\epsilon_2-i\gamma_2/2}\frac{\epsilon_1+\epsilon_2-i\eta}{\epsilon_1+\epsilon_2-i\gamma/2} \\ \nonumber
-T_{DL}^2\frac{1}{\epsilon_{\bf q}+\epsilon_1-i\gamma_1/2}\frac{\epsilon_1+\epsilon_2-i\eta}{\epsilon_1+\epsilon_2-i\gamma/2}.
\end{eqnarray}
This finally leads to
\begin{eqnarray}
\label{T}
\langle pq|T'|DD\rangle = -T_{DL}^2 \frac{\epsilon_1+\epsilon_2-i\eta}
{\epsilon_1+\epsilon_2-i\gamma/2}\times \\ \nonumber
\frac{\epsilon_{\bf p}+\epsilon_{\bf q}+\epsilon_1+\epsilon_2-i\gamma/2}{(\epsilon_{\bf p}+\epsilon_2-i\gamma_2/2)(\epsilon_{\bf q}+\epsilon_1-i\gamma_1/2)}.
\end{eqnarray}
When neglecting the resistances of the leads that contain the capacitors $C_1$ and $C_2$, one can
exploit that $\epsilon_{\bf q}+\epsilon_{\bf p}=2\mu_S=0$ and recover the result from Ref. \onlinecite{Recher}
\begin{eqnarray}
\langle pq|T'|DD\rangle = -T_{DL}^2 (\epsilon_1+\epsilon_2-i\eta)\times
\\ \nonumber
\frac{1}{(\epsilon_{\bf p}+\epsilon_2-i\gamma_2/2)(\epsilon_{\bf q}+\epsilon_1-i\gamma_1/2)}.
\end{eqnarray}

\section{Calculation of $\langle [d_{2\downarrow} d_{1\uparrow}
\pm d_{2\uparrow} d_{1\downarrow}]T"\rangle$}

Here, we would like to compute $(1/\sqrt{2})\langle [d_{2\downarrow} d_{1\uparrow}
\pm d_{2\uparrow} d_{1\downarrow}]T"\rangle$ where the abbreviation $\langle ... \rangle$ stands for
$\langle i|...| i \rangle$. This part is formally equivalent to
\begin{eqnarray}
\langle DD|T''|i\rangle = \frac{1}{\sqrt{2}}\langle i | [d_{2\downarrow} d_{1\uparrow} - d_{2\uparrow} d_{1\downarrow}]
\times \\ \nonumber
\frac{1}{i\eta-H_0} H_{SD_1} \frac{1}{i\eta-H_0} H_{SD_2} |i\rangle.
\end{eqnarray}
We can already evaluate $\langle i | [d_{2\downarrow} d_{1\uparrow} - d_{2\uparrow} d_{1\downarrow}]\frac{1}{i\eta-H_0}=\langle i | [d_{2\downarrow} d_{1\uparrow} - d_{2\uparrow} d_{1\downarrow}]\frac{1}{i\eta-\epsilon_1-\epsilon_2}$ leading to
\begin{eqnarray}
\hskip -0.6cm \langle DD|T''|i\rangle = \frac{1}{\sqrt{2}}\frac{1}{i\eta-\epsilon_1-\epsilon_2} 
\times \\ \nonumber
\langle i | [d_{2\downarrow} d_{1\uparrow} - d_{2\uparrow} d_{1\downarrow}]H_{SD_1} \frac{1}{i\eta-H_0} H_{SD_2} |i\rangle.
\end{eqnarray}
Now, following Ref. \onlinecite{Recher}, we insert a complete set of single-particle (virtual) states
\begin{equation}
1=\sum_{l{\bf k}\sigma} \gamma^{\dagger}_{{\bf k}\sigma} d^{\dagger}_{l-\sigma}|i\rangle\langle i|d_{l-\sigma}\gamma_{{\bf k}\sigma},
\end{equation}
such that
\begin{eqnarray}
\langle DD|T''|i\rangle = \frac{1}{\sqrt{2}}\frac{1}{i\eta-\epsilon_1-\epsilon_2}
\langle i | [d_{2\downarrow} d_{1\uparrow} - d_{2\uparrow} d_{1\downarrow}]
H_{SD_1}  \hskip 0.3cm \\ \nonumber
\times\sum_{l{\bf k}\sigma} \gamma^{\dagger}_{{\bf k}\sigma} d^{\dagger}_{l-\sigma}|i\rangle\langle i|d_{l-\sigma}\gamma_{{\bf k}\sigma}
\frac{1}{i\eta-H_0} H_{SD_2} |i\rangle.
\end{eqnarray}
Now we can use $\langle i|d_{l-\sigma}\gamma_{{\bf k}\sigma}
\frac{1}{i\eta-H_0}=\langle i|d_{l-\sigma}\gamma_{{\bf k}\sigma}
\frac{1}{i\eta-E_{\bf k}-\epsilon_l}$; $E_{\bf k}$ being the energy of a BCS quasiparticle. Moreover, considering
that the dots are at resonance we can approximate $i\eta-E_{\bf k}-\epsilon_l\approx - E_{\bf k}$ and
obtain
\begin{eqnarray}
\langle DD|T''|i\rangle = \frac{1}{\sqrt{2}}\frac{1}{i\eta-\epsilon_1-\epsilon_2}
\langle i | [d_{2\downarrow} d_{1\uparrow} - d_{2\uparrow} d_{1\downarrow}]H_{SD_1}
\hskip 0.5cm
\\ \nonumber
\times \sum_{l{\bf k}\sigma} \frac{1}{-E_{\bf k}} \gamma^{\dagger}_{{\bf k}\sigma} d^{\dagger}_{l-\sigma}|i\rangle\langle i|d_{l-\sigma}\gamma_{{\bf k}\sigma} H_{SD_2} |i\rangle\\ \nonumber
= \frac{1}{\sqrt{2}}\frac{T_{SD}^2}{-i\eta+\epsilon_1+\epsilon_2}
\langle i | [d_{2\downarrow} d_{1\uparrow} - d_{2\uparrow} d_{1\downarrow}]
\sum_{l{\bf k}\sigma}\sum_{\sigma'}\sum_{\sigma''}\times
\\ \nonumber
\frac{1}{E_{\bf k}}
d^{\dagger}_{1\sigma'}\Psi_{\sigma'}({\bf r}_1)
\gamma^{\dagger}_{{\bf k}\sigma} d^{\dagger}_{l-\sigma}|i\rangle\langle i|d_{l-\sigma}\gamma_{{\bf k}\sigma} 
d_{2\sigma''}^{\dagger}\Psi_{\sigma''}({\bf r}_2)|i\rangle.
\end{eqnarray}
The terms which survive are those with $l=2$ and $\sigma'=-\sigma''=\sigma$ ($-\sigma$ is the spin polarization opposite to $\sigma$):
\begin{eqnarray}
|\langle DD|T''|i\rangle| = \frac{1}{\sqrt{2}}\frac{T_{SD}^2}{-i\eta+\epsilon_1+\epsilon_2}
\langle i | [d_{2\downarrow} d_{1\uparrow} - d_{2\uparrow} d_{1\downarrow}] \hskip 0.5cm \\ \nonumber
\times\sum_{{\bf k}\sigma}\frac{1}{E_{\bf k}}
d^{\dagger}_{1\sigma}\Psi_{\sigma}({\bf r}_1)
\gamma^{\dagger}_{{\bf k}\sigma} d^{\dagger}_{2-\sigma}|i\rangle\langle i|\gamma_{{\bf k}\sigma}\Psi_{-\sigma}({\bf r}_2) d_{2-\sigma} d^{\dagger}_{2-\sigma}|i\rangle.
\end{eqnarray}
Now we can develop $\Psi_{-\sigma}(r_2)=\sum_{{\bf k}'} e^{i{\bf k}' r_1} c_{{\bf k}'-\sigma}$ as well as
$\Psi_{\sigma}(r_1)=\sum_{{\bf k}''} e^{i{\bf k}'' r_1} c_{{\bf k}''\sigma}$, and exploit the precious equalities
$\langle i| \gamma_{{\bf k}\sigma} c_{{\bf k}'-\sigma}=\langle i| v_{\bf k} \delta({\bf k+k}')\epsilon_{\sigma}$ and $\langle i| c_{{\bf k}''\sigma} \gamma^{\dagger}_{{\bf k}\sigma} = \langle i| u_{\bf k} \delta({\bf k-k}'')$ which have been demonstrated in Appendix A. Note that in agreement with the BCS theory we satisfy
${\bf k}' = - {\bf k}''$. Hence we converge to
\begin{eqnarray}
\label{intric}
|\langle DD|T''|i\rangle| = \frac{1}{\sqrt{2}}\frac{T_{SD}^2}{-i\eta+\epsilon_1+\epsilon_2}
\langle i | [d_{2\downarrow} d_{1\uparrow} - d_{2\uparrow} d_{1\downarrow}] \hskip 0.3cm \\ \nonumber
\times\sum_{{\bf k}\sigma}\frac{u_{\bf k} v_{\bf k}}{E_k} e^{i{\bf k}({\bf r}_1-{\bf r}_2)} \epsilon_\sigma 
d^{\dagger}_{1\sigma}d^{\dagger}_{2-\sigma}|i\rangle.
\end{eqnarray}
Now we can resort to the important equality
\begin{equation}
\label{ent}
\sum_{\sigma} \langle i | [d_{2\downarrow} d_{1\uparrow} - d_{2\uparrow} d_{1\downarrow}] \epsilon_\sigma 
d^{\dagger}_{1\sigma}d^{\dagger}_{2-\sigma}|i\rangle =  2,
\end{equation}
then leading to
\begin{equation}
|\langle DD|T''|i\rangle| = \frac{\sqrt{2}T_{SD}^2}{-i\eta+\epsilon_1+\epsilon_2}\sum_{{\bf k}}
\frac{u_{\bf k} v_{\bf k}}{E_{\bf {k}}} e^{i{\bf k}({\bf r}_1-{\bf r}_2)}.
\end{equation}
In fact, another term which consists of exchanging the roles of $H_{SD_1}$
and $H_{SD_2}$ or ${\bf r}_1$ and ${\bf r}_2$ in Eq. (C1) should be also included (see Eq. (12)). This means that
\begin{eqnarray}
|\langle DD|T''|i\rangle| = \frac{\sqrt{2}T_{SD}^2}{-i\eta+\epsilon_1+\epsilon_2}\sum_{{\bf k}}
\\ \times
\frac{u_{\bf k} v_{\bf k}}{E_{\bf k}} \left(e^{i{\bf k}({\bf r}_1-{\bf r}_2)}+e^{-i{\bf k}({\bf r}_1-{\bf r}_2)}\right),
\end{eqnarray}
and finally
\begin{equation}
|\langle DD|T''|i\rangle| = \frac{2\sqrt{2}T_{SD}^2}{-i\eta+\epsilon_1+\epsilon_2}\sum_{{\bf k}}
\frac{u_{\bf k} v_{\bf k}}{E_{\bf k}} 
\cos({\bf k}\delta {\bf r}).
\end{equation}
Now, we can exploit $u_{\bf k} v_{\bf k}=\frac{1}{2 E_{\bf k}}\sqrt{E_{\bf k}^2-\xi_{\bf k}^2}=
\Delta/(2E_{\bf k})$:
\begin{equation}
|\langle DD|T''|i\rangle| = \frac{2\sqrt{2}T_{SD}^2}{-i\eta+\epsilon_1+\epsilon_2}\sum_{{\bf k}}\frac{\Delta}{2 E_{\bf k}^2}\cos({\bf k}.\delta {\bf r}).
\end{equation}
Here, we can assume that $|{\bf k}|\approx k_F$ and we note ${\bf k}_F.\delta{\bf r}=k_F\delta r\sin\theta$ where $\delta r=|\delta{\bf r}|$ such that
 \begin{equation}
|\langle DD|T''|i\rangle| \sim \frac{2\sqrt{2}T_{SD}^2}{-i\eta+\epsilon_1+\epsilon_2}\frac{1}{2\Delta}\int_0^{\pi} d\theta \cos(k_F\delta r\sin\theta).
\end{equation}
We recover the crossed Andreev contribution of Ref. \onlinecite{Recher}
\begin{equation}
|\langle DD|T''|i\rangle| \approx \frac{1}{\sqrt{2}}\frac{\gamma_S}{-i\eta+\epsilon_1+\epsilon_2}\frac{\sin({k}_F\delta {r})}{{k}_F \delta { r}},
\end{equation}
which is still valid in the presence of voltage noise.

\section{Current calculations from crossed Andreev reflection}

\subsection{No dissipation}

Without dissipation, from Appendices B and C we get
\begin{eqnarray}
I &=& \frac{e}{\hbar} \sum_{{\bf p},{\bf q}} |T_{DL}|^4
2\pi \gamma_S^2\left(\frac{\sin({k}_F\delta {r})}{{k}_F \delta { r}}\right)^2 
\\ \nonumber
&\times& 
\left| \frac{1}{(\epsilon_{\bf p}+\epsilon_2-i\gamma_2/2)(\epsilon_{\bf q}+\epsilon_1-i\gamma_1/2)}
\right|^2 \delta(\epsilon_{\bf p}+\epsilon_{\bf q}).
\end{eqnarray}
Now, when $\epsilon_{\bf p}+\epsilon_{\bf q}=0$ we can rewrite
\begin{eqnarray}
& &\frac{1}{(\epsilon_{\bf p}+\epsilon_2-i\gamma_2/2)(\epsilon_{\bf q}+\epsilon_1-i\gamma_1/2)}
=  \\ \nonumber
& &\frac{1}{\epsilon_1+\epsilon_2-i\gamma/2} \left(\frac{1}{\epsilon_{\bf p}+\epsilon_2-i\gamma_2/2}+
\frac{1}{-\epsilon_{\bf p}+\epsilon_1-i\gamma_1/2}\right),
\end{eqnarray}
and hence the current turns into
\begin{eqnarray}
I = \sum_{{\bf p}} \frac{e}{2\hbar} |T_{DL}|^2 
\frac{\gamma\gamma_S^2}{(\epsilon_1+\epsilon_2)^2+\gamma^2/4}\left(\frac{\sin({k}_F\delta {r})}{{k}_F \delta { r}}\right)^2 
\\ \nonumber
\times  \left|\frac{1}{\epsilon_{\bf p}+\epsilon_2-i\gamma_2/2}+
\frac{1}{-\epsilon_{\bf p}+\epsilon_1-i\gamma_1/2}\right|^2.
\end{eqnarray}
Owing to the Breit-Wigner resonance, we can eventually simplify $\epsilon_{\bf p}\sim \epsilon_c \sim \epsilon_1$ and thus this results in
\begin{eqnarray}
 \left|\frac{1}{\epsilon_{\bf p}+\epsilon_2-i\gamma_2/2}+
\frac{1}{-\epsilon_{\bf p}+\epsilon_1-i\gamma_1/2}\right|^2 = \frac{4\pi }{\gamma_2} \delta(\epsilon_{\bf p}-\epsilon_1), \hskip 0.3cm
\end{eqnarray}
which results in:
\begin{eqnarray}
I = \frac{e}{\hbar}\frac{\gamma\gamma_S^2}{(\epsilon_1+\epsilon_2)^2+\gamma^2/4}\left(\frac{\sin({k}_F\delta {r})}{{k}_F \delta { r}}\right)^2.
\end{eqnarray}
This is the formula in Ref. \onlinecite{Recher}. Exactly at the resonance condition for the dots where $\epsilon_1=-\epsilon_2=0$ the crossed Andreev reflection gives 
\begin{eqnarray}
I = \frac{4e\gamma_S^2}{\hbar \gamma}\left(\frac{\sin({k}_F\delta {r})}{{k}_F \delta { r}}\right)^2.
\end{eqnarray}
It is important to keep in mind that one requires the electrons residing simultaneously on the dots to be in the {\it singlet} state configuration so that results of Appendix C can be safely applied\cite{Recher}. This is well satisfied when the distance between the dots $<\xi$. Indeed, the current carried by the $S_z=0$ configuration of the triplet state on the dots would be zero because
\begin{equation}
\label{ent}
\sum_{\sigma} \langle i | [d_{2\downarrow} d_{1\uparrow} + d_{2\uparrow} d_{1\downarrow}] \epsilon_\sigma 
d^{\dagger}_{1\sigma}d^{\dagger}_{2-\sigma}|i\rangle =  0.
\end{equation}
In the same spirit, to be non-zero Eq. (C7) assumes that the two injected electrons have anti-parallel spin configurations $\sigma''=-\sigma'$ and opposite momenta ${\bf k}''=-{\bf k}'$ which is well-satisfied for $\delta r<\xi$. Therefore we can rewrite
\begin{eqnarray}
\label{current}
I[R_1=0] = \frac{4e\gamma_S^2}{\hbar \gamma}\left(\frac{\sin({k}_F\delta {r})}{{k}_F \delta { r}}\right)^2 e^{-2\delta r/\pi \xi}.
\end{eqnarray}

\subsection{Weak dissipation on dot 1}

For a weak resistance $R_1$ so that $\alpha_1=R_K/R_1\gg 1$ the EPR current can be approximated as
\begin{eqnarray}
\hskip -0.5cm I &\sim& \frac{e}{\hbar} \sum_{{\bf p},{\bf q}} |T_{DL}|^4
2\pi \gamma_S^2\left(\frac{\sin({k}_F\delta {r})}{{k}_F \delta { r}}\right)^2 P_1(2\mu_S-\epsilon_{\bf p}-\epsilon_{\bf q}) \hskip 0.8cm
\\ \nonumber
&\times& 
\left| \frac{1}{(\epsilon_{\bf p}+\epsilon_2-i\gamma_2/2)(\epsilon_{\bf q}+\epsilon_1-i\gamma_1/2)}
\right|^2.
\end{eqnarray}
Since we consider that the dots are close to resonance, we have replaced $P_1(2\mu_S-\epsilon_1-\epsilon_2)\sim \delta(2\mu_S-\epsilon_1-\epsilon_2)$ where the function $P_1(E)$ (which yields a blatant singularity at $E=0$) has been precisely defined in Eq. (\ref{P1}).  Now, we concentrate mainly on the effect of the quantum noise on the two particle Breit-Wigner resonance between the dots and the leads. For large $\alpha_1$ again the function $P_1(E)$ has unambiguously a pronounced singularity at $E=0$ implying that $\epsilon_{\bf p} +\epsilon_{\bf q}\rightarrow 0$ in the tunneling process between the dots and the leads. Using Eq. (D4) this allows us to approximate
\begin{eqnarray}
I &\sim& \frac{e}{\hbar} \sum_{{\bf p},{\bf q}}
 \frac{2\pi |T_{DL}|^4\gamma_S^2}{(\epsilon_1+\epsilon_2)^2+\gamma^2/4}
\left(\frac{\sin({k}_F\delta {r})}{{k}_F \delta { r}}\right)^2 
\\ \nonumber
&\times&  \frac{4\pi }{\gamma_2} \delta(\epsilon_{\bf p}-\epsilon_1)P_1(2\mu_S-\epsilon_{\bf p}-\epsilon_{\bf q}).
\end{eqnarray}
We then converge to
\begin{eqnarray}
I \sim \frac{e}{\hbar}\frac{\gamma\gamma_S^2}{(\epsilon_1+\epsilon_2)^2+\gamma^2/4}\left(\frac{\sin({k}_F\delta {r})}{{k}_F \delta { r}}\right)^2 \\ \nonumber
\times\int_{\mu_l}^{\epsilon_c\sim \epsilon_2} d\epsilon_{\bf q} P_1(2\mu_S-\epsilon_1-\epsilon_{\bf q}).
\end{eqnarray}
We have used the fact that the dots are close to the resonance condition $\epsilon_1+\epsilon_2=0$. Using
Eq. (\ref{P1}) we obtain
\begin{eqnarray}
I \sim \frac{e}{\hbar}\frac{\gamma\gamma_S^2}{(\epsilon_1+\epsilon_2)^2+\gamma^2/4}\left(\frac{\sin({k}_F\delta {r})}{{k}_F \delta { r}}\right)^2 
\\ \nonumber
\frac{\exp(-2\gamma_e/\alpha_1)}{\Gamma(1+2/\alpha_1)}
\left(\frac{\pi}{\alpha_1}\right)^{2/\alpha_1}\left(\frac{2\mu}{E_{c1}}\right)^{2/\alpha_1}.
\end{eqnarray}
Hence we can summarize
\begin{eqnarray}
I[\alpha_1\gg 1] \sim {I[R_1=0]} \frac{\exp(-2\gamma_e/\alpha_1)}{\Gamma(1+2/\alpha_1)}
\left(\frac{2\mu \pi}{\alpha_1 E_{c1}}\right)^{2/\alpha_1}. \hskip 0.6cm
\end{eqnarray}

\subsection{Two independent environments}

In the case of two symmetric and moderate environments such that $R_1,R_2\ll R_K$, we get
\begin{eqnarray}
I \sim \frac{e}{\hbar}\frac{\gamma\gamma_S^2}{(\epsilon_1+\epsilon_2)^2+\gamma^2/4}\left(\frac{\sin({k}_F\delta {r})}{{k}_F \delta { r}}\right)^2 \\ \nonumber
\times\int_{\mu_l}^{\epsilon_c\sim \epsilon_2} d\epsilon_{\bf q} P_{12}(2\mu_S-\epsilon_1-\epsilon_{\bf q}),
\end{eqnarray}
where the function $P_{12}(E)$ has been defined in Eq. (37).  Assuming that $E_{c1}\sim E_{c2}$ this leads to
\begin{eqnarray}
I[\alpha_l\gg 1] \sim {I[R_1=0]} \frac{\exp(-2\gamma_e/\alpha)}{\Gamma(1+2/\alpha)}
\left(\frac{2\mu \pi}{\alpha E_{c1}}\right)^{2/\alpha}. \hskip 0.6cm
\end{eqnarray}

\section{Current due to direct Andreev processes}

Using Eq. (\ref{amp}), for weak resistances $\alpha_l\gg 1$ we find that the current $I'=I^{(I)} + I^{(II)}$ 
due to direct Andreev processes (when the two electrons tunnel to the same dot) can be easily valued leading to
\begin{eqnarray}
I'  &\sim& \frac{2e}{\hbar} \sum_{{\bf p}, {\bf p}'} \sum_l  {\pi}\frac{\gamma_S^2 |T_{DL}|^4}{{\cal E}^2}
P_l(2\mu_S - \epsilon_{\bf p} - \epsilon_{{\bf p}'}) 
\\ \nonumber
&\times& \left| \frac{1}{\epsilon_{\bf p} +\epsilon_l -i\gamma_l/2} + \frac{1}{\epsilon_{{\bf p}'} +\epsilon_l -i\gamma_l/2} \right|^2. \\ \nonumber
\end{eqnarray}
We have approximated $P_l(2\mu_S-\epsilon_{{\bf p}''}-\epsilon_l) \sim \delta(2\mu_S-\epsilon_{{\bf p}''}-\epsilon_l)$ and we have explicitly introduced ${\cal E}^{-1} = 1/(\pi \Delta) + 1/U$. 
In the absence of noise, the energy conservation implies $\epsilon_{\bf p} + \epsilon_{{\bf p}'}=0\sim \epsilon_l$ and therefore one can exploit
\begin{eqnarray}
\left|\frac{1}{\epsilon_{\bf p} +\epsilon_l -i\gamma_l/2} + \frac{1}{\epsilon_{{\bf p}'} +\epsilon_l -i\gamma_l/2} \right|^2 = \frac{4\pi}{\gamma_l} \delta(\epsilon_{\bf p} -\epsilon_l), \hskip 0.4cm
\end{eqnarray}
resulting in 
\begin{eqnarray}
I' [R_l=0] &=& \frac{4e}{\hbar} \sum_{{\bf p}, {\bf p}'} {4\pi^2}\frac{\gamma_S^2 |T_{DL}|^4}{\gamma_l {\cal E}^2}
\delta(\epsilon_{\bf p} + \epsilon_{{\bf p}'}) \delta(\epsilon_{\bf p} -\epsilon_l) \hskip 0.8cm \\ \nonumber
&=& \frac{4e}{\hbar} {4\pi^2 \nu_l^2}\frac{\gamma_S^2 |T_{DL}|^4}{\gamma_l {\cal E}^2} = \frac{2e \gamma_S^2 \gamma}{\hbar {\cal E}^2}.
\end{eqnarray}
Again, we reproduce the result announced in Ref. \onlinecite{Recher}. In the limit of weak ohmic
resistors $R_1$ and $R_2$ we estimate
\begin{eqnarray}
I' [\alpha_l \gg 1]\sim \frac{I'[R_l=0]}{2} \sum_l \int_{\mu_l}^{\epsilon_c\sim \epsilon_l} d\epsilon_{\bf q} P_l(2\mu_S-\epsilon_l - \epsilon_{\bf q}) \hskip 0.4cm \\ \nonumber
= \frac{I'[R_l=0]}{2} \sum_l \frac{\exp(-2\gamma_e/\alpha_l)}{\Gamma(1+2/\alpha_l)} \left(\frac{2\mu \pi}{\alpha_l E_{cl}}\right)^{2/\alpha_l}.  
\end{eqnarray}
Note that for the most probable situation of symmetric environments, the suppression factor $(2\mu/E_{cl})^{2/\alpha_l}$ is less considerable than that for the EPR-pair current $I$. We argue that this is well justified in this setup because  when two spin-entangled electrons tunnel onto the same dot the Coulomb blockade forbids charge-2e transport; this results in a single-particle Breit-Wigner resonance between, say, dot $l$, and lead $l$
which is less affected by the baths than the two-particle
Breit-Wigner resonance.


\begin{thebibliography}{10}
\bibitem{note3}
We have $\hat{Q}_{nl}=\sqrt{2}\int_0^1 dx \cos\left(\pi n x +\pi x/2\right) \hat{Q}_l(x)$
and $\hat{\phi}_{nl}=\sqrt{2}\int_0^1 dx \cos\left(\pi n x +\pi x/2\right) \hat{\phi}_l(x)$.
\bibitem{noteC}
Note that we have reversed the order of some bath operators. Since all the operators are taken at equal
time, we have assumed that the ``photons'' are completely emitted only when the two electrons reside in the leads.
\bibitem{note2}
We extract $W_{DD,f}=\left|\langle pq| T' |DD\rangle\right|^2 \int_{-\infty}^{+\infty} \frac{dt}{2\pi\hbar} e^{i(\epsilon_f-\epsilon_1-\epsilon_2)t/\hbar}\times
\langle D_B| e^{-i\delta\phi_1(t)} e^{i\delta\phi_1(0)} |D_B \rangle$. We use $|D_B\rangle = e^{-i\delta\phi_1(0)} |i_B\rangle$ and $P_1(E)$ is 
 a real function meaning that $P_1(E)^* = P_1(E)$.
\end{thebibliography}
\end{document}